\documentclass[%
 reprint,
superscriptaddress,
 amsmath,amssymb,
 aps,
 prl,
]{revtex4-2}

\usepackage{graphicx}% Include figure files
\usepackage{dcolumn}% Align table columns on decimal point
\usepackage{bm}% bold math
\usepackage{hyperref}
\usepackage{makecell}
\usepackage{multirow}
\usepackage{bigstrut}

\usepackage{xcolor}

\begin{document}

\preprint{APS/123-QED}

\title{Itinerant magnetism in the  triangular lattice Hubbard model at half-doping: application to twisted transition-metal dichalcogenides}

\author{Yuchi He}
\email{yuchi.he@physics.ox.ac.uk}
	\affiliation{Rudolf Peierls Centre for Theoretical Physics, Parks Road, Oxford, OX1 3PU, United Kingdom}
 \author{Roman Rausch}
  \affiliation{Technische Universität Braunschweig, Institut für Mathematische Physik,
  Mendelssohnstraße 3, 38106 Braunschweig, Germany}
  \author{Matthias Peschke}
	\affiliation{ Institute for Theoretical Physics Amsterdam and Delta Institute for Theoretical
Physics, University of Amsterdam, Science Park 904, 1098 XH Amsterdam, The
Netherlands}
	\affiliation{I. Institute of Theoretical Physics, University of Hamburg, Notkestraße 9, 22607
Hamburg, Germany}
\author{Christoph~Karrasch}
  \affiliation{Technische Universität Braunschweig, Institut für Mathematische Physik,
  Mendelssohnstraße 3, 38106 Braunschweig, Germany}
\author{Philippe Corboz}
	\affiliation{Institute for Theoretical Physics Amsterdam and Delta Institute for Theoretical
Physics, University of Amsterdam, Science Park 904, 1098 XH Amsterdam, The
Netherlands}
	\author{Nick Bultinck}
	\affiliation{Department of Physics and Astronomy, Ghent University, Krijgslaan 281, 9000 Gent, Belgium}
 \author{S.A. Parameswaran}
 \affiliation{Rudolf Peierls Centre for Theoretical Physics, Parks Road, Oxford, OX1 3PU, UK}

\date{\today}

\begin{abstract}
We use unrestricted Hartree-Fock, density matrix renormalization group, and variational projected entangled-pair state calculations to investigate the ground-state phase diagram of the triangular-lattice Hubbard model at ``half doping'' relative to single occupancy, i.e., at a
filling of $(1\pm \frac{1}{2})$ electrons per site. The electron-doped case has a nested Fermi surface in the noninteracting limit, and hence a weak-coupling instability towards density-wave orders whose wave vectors are determined by Fermi-surface nesting conditions. We find that at moderate-to-strong interaction strengths, other spatially modulated orders arise, with wave vectors distinct from the nesting vectors. In particular, we identify a series of closely competing itinerant long-wavelength magnetically ordered states, yielding to uniform ferromagnetic order at the largest interaction strengths. For half-hole doping and a similar range of interaction strengths, our data indicate that magnetic orders are most likely absent.
\end{abstract}

\maketitle

The triangular-lattice Hubbard model plays a paradigmatic role in studying the interplay between electronic interactions and geometric frustration.
At half filling ($\nu=1$ electron per site),  double-occupancy is suppressed by Hubbard repulsion; electron spins are then the dominant degrees of freedom, but their ordering is frustrated by the  non-bipartite nature of the triangular lattice~\cite{WannierIsingtriangular}--- a scenario believed to favor the formation of 
quantum spin liquids. On doping away from half filling, strong-coupling expansion yields a picture of geometrically frustrated magnetic moments coupled to itinerant electrons. The existence and nature of magnetic order in this setting remain a challenging question.

Experimentally, layered materials with triangular lattice structure~\cite{Shimizu2003,Takada2003, Hasan2004,Foo2004,Lee06,Zhou11,Zvyagin2019} provide a natural platform for realizing the triangular-lattice Hubbard model (TLHM). More recently, effective extended TLHMs have also been constructed~\cite{Wu} to describe electrons in homo- and hetero-bilayer transition metal dichalcogenide (TMD) moir\'e materials~\cite{Zhang2017,Pan2018,Wang2020}; experimental evidence of strong correlations has recently been reported~\cite{Wang2020,Regan2020,Tang2020,Huang2021,Jin2021,Li2021,Xu2022,Tang2023,ciorciaro2023kinetic}. The TLHM has also been realized in a more controlled setting in cold-atom experiments~\cite{Xu2023,lebrat2023observation, prichard2023directly}.

Most existing theoretical efforts to establish the ground-state phase diagram of the TLHM and related or extended models~\cite{SinghHuse,ANDP, KineticAFM,Li2dtensor,ZhuTriangular,JiangTriangular, PanWu1, APLM, PengPDW,WietekWang, Zhu,PhysRevB.106.235135,chen2023singlet,zegrodnik2023mixed} have studied half filling or a small filling range
straddling it. A smaller body of work~\cite{ANDP,Honerkamp,MartinBatista,li2010metalinsulator,YeRan,PasrijaKumar,PanWu2,ZangWang,PhysRevB.99.195120,SchererKennes,Yao,DMFTTLHM,Xu2023,potasz2023itinerant}  has studied the case of ``half electron doping"  ($\nu=3/2$), mostly motivated by experiments described above. Theoretical efforts have also built on the resemblance of the noninteracting problem to that of graphene doped to a filling of $1\pm 1/4$ electron per site~\cite{Li_2012,WangXiang,NandkishoreLevitov, NandkishorChern,JiangMesaros}, namely, a nested Fermi surface whose associated Van Hove singularity signals a weak-coupling instability to broken-symmetry order at the nesting wave vectors. In the half-electron-doped TLHM case, the  weak-coupling order is predicted to be an unusual magnetic insulator with tetrahedral spin order~\cite{MartinBatista,AkagiMotome}. The possibility of realizing this exotic broken-symmetry state, and its potential to stabilize  chiral superconductivity, have stimulated much experimental and theoretical work.

However, two recent experimental studies of twisted TMDs~\cite{Wang2020,Li2021} --- theoretically modeled as the single-band TLHM model or simple extensions --- find no evidence for insulating states at 1/2 electron doping. This motivates our study of the TLHM ground-state phase diagram at larger interaction strengths, where the weak-coupling assumption is no longer valid.
We will also consider the quantum phases at 1/2 hole doping, demonstrating the clear particle-hole asymmetry of the TLHM.

\begin{figure}
\begin{center}
\includegraphics[width=1\columnwidth]{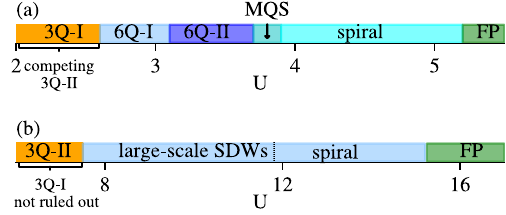}
\end{center}
\caption{Ground-state phase diagram of the TLHM as a function of $U$ at 1/2 electron doping. The magnetic orders of the different phases are illustrated in Table.~\ref{tbl:HForders}. The abbreviation $n$Q stands for spin-density waves with $n$ momentum component. The I stands for noncoplanar, and II stands for collinear. MQS:multi-Q noncoplanar stripe. FP: full spin polarization. (a) Phase diagram obtained from unrestricted Hartree-Fock simulations.  (b) Phase diagram inferred from tensor network calculations. The obtained large-scale SDWs include collinear orders akin to 6Q-II as well as (coplanar) spiral order.
}\label{fig:PD}
\end{figure}

In the absence of a controlled weak-coupling calculation, we have deployed a variety of numerical variational methods to study the TLHM: (a)~unrestricted self-consistent Hartree-Fock~\cite{ODA} (HF), (b)~the density matrix renormalization group~\cite{White,SingleSite} (DMRG), and (c) infinite  projected entangled-pair states~\cite{verstraete2004,nishio2004tensor,CTM,fPEPS,iePEPS,simpleuupdate,VPEPS,PhysRevB.94.155123,ADPEPS,AD2} (PEPS).
At 1/2 electron doping, we find that beyond a critical interaction strength the proposed insulating spin-tetrahedral state gives way to other magnetic orders with different wave vectors. 
This is in contrast to the half-filled square-lattice Hubbard model, where the spin-density-wave (SDW) momentum ($\pi,\pi)$ is equal to the nesting vector for all interaction strengths. Starting from moderate interaction strengths, our numerics reveal a rich phase diagram hosting a series of different large-scale magnetically ordered metallic states, see Fig.~\ref{fig:PD}. For weaker interactions, we also find a metallic collinear-spin state which competes with the insulating spin-tetrahedral state. At the largest interaction strengths, uniform ferromagnetism emerges.  For 1/2 hole doping, we find no evidence for magnetism up to reasonably strong interactions.

\paragraph{Hamiltonian.}
We consider the following  Hamiltonian on the triangular lattice with nearest-neighbor ($\langle i,j\rangle$) hopping and density-density interactions:
\begin{align}\label{eq:Hamiltonian}
\!\!\!\! H=-\!\!\!\sum_{\langle i,j \rangle, \sigma}\!\! c_{i\sigma}^{\dagger}c^{\phantom\dagger}_{j\sigma}\!+ U\sum_{i}n_{i\uparrow}n_{i\downarrow}  
\!+\! V \sum_{i \neq j} f(r_{ij}) n_i n_j,
\end{align}
where $c_{i\sigma}$ ($c^{\dagger}_{i\sigma}$) is the annihilation (creation) operator of electron at site $i$ with spin index $\sigma=\uparrow, \downarrow$; and $n_{i\sigma}= c^{\dagger}_{i\sigma}c_{i\sigma}$ ($n_{i}=\sum_\sigma n_{i\sigma}$)  is the density operator of spin-$\sigma$ electrons (and the total density); $r_{ij}$ is the distance between site $i$ and $j$ in units of the lattice constant.  
We are primarily interested in the repulsive Hubbard model $U>0$, $V=0$, and will comment on the effects of long-range interaction $Vf(r)$.
We focus on the quantum phases at 1/2 electron doping with respect to half filling, i.e., $\overline{\langle n_i \rangle}=1+\frac{1}{2}=\frac{3}{2}$, where the average (overline) is over all sites. We will also comment on the 1/2 hole-doping case $\overline{\langle n_i \rangle}=\frac{1}{2}$.

\setlength{\bigstrutjot}{10pt}

\begin{table}[h!]
  \centering
  \begin{tabular}{ | m{1.0cm} | m{2.5cm} | m{3.3cm} | }
    \hline
    Names &  Momentum $\bm{Q}$ & Magnetic unit cell \\ \hline
    
    &
    \multirow[c]{6}{*}[0in]{
      \begin{minipage}[t]{0.\textwidth}
      \vspace{-2 mm}
      \includegraphics{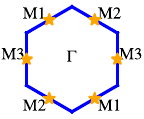}
    \end{minipage} 
    }
    & 
     \multirow[c]{3}{*}[0in]{
      \begin{minipage}[t]{0.\textwidth}
      \vspace{-12 mm}
      \hspace*{-5 mm}
      \includegraphics{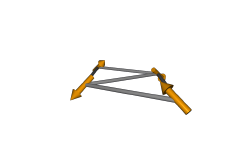}
    \end{minipage} 
    }
    \\ 
    3Q-I
    &
    &
    \\
    
    &
    &
    \\ \cline{1-1} \cline{3-3} 
    
    &
    &
    \multirow[c]{3}{*}[0in]{
      \begin{minipage}[t]{0.\textwidth}
      \vspace{-8 mm}
      \includegraphics{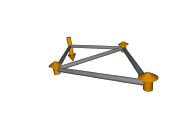}
    \end{minipage} 
    } \\ 
    3Q-II
    &
    &
    \\ 
    
    &
    &
    \\ \hline

    &
    \multirow[c]{6}{*}[0in]{
      \begin{minipage}[t]{0.\textwidth}
      \vspace{-2 mm}
      \includegraphics{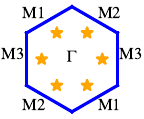}
    \end{minipage} 
    }
    & 
    \\ 
    6Q-I
    &
    &
    Incommensurate \newline noncoplanar order
    \\
    
    &
    &
    \\ \cline{1-1} \cline{3-3} 
    
    &
    &
    \multirow[c]{3}{*}[0in]{
      \begin{minipage}[t]{0.\textwidth}
      \vspace{-8.5 mm}
      \hspace*{3 mm}
      \includegraphics{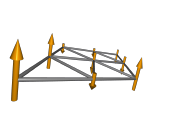}
    \end{minipage} 
    } \\ 
    6Q-II
    &
    &
    \\ 
    
    &
    &
    \\ \hline

    &
    \multirow[c]{6}{*}[0in]{
      \begin{minipage}[t]{0.\textwidth}
      \vspace{-2 mm}
      \includegraphics{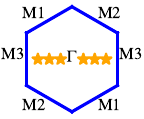}
    \end{minipage} 
    }
    & 
     \multirow[c]{3}{*}[0in]{
      \begin{minipage}[t]{0.\textwidth}
      \vspace{-16 mm}
      \hspace*{-3 mm}
      \includegraphics{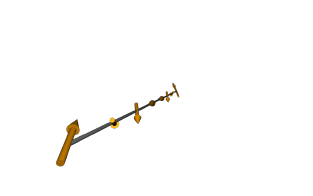}
    \end{minipage} 
    }
    \\ 
    
    &
    &
    \\
    MQS
    &
    &
    \\ 
    
    &
    &
    \\ 
    
    &
    &
    Commensurate \newline noncoplanar order
    \\ 
    
    &
    &
    \\ \hline

    &
    \multirow[c]{6}{*}[0in]{
      \begin{minipage}[t]{0.\textwidth}
      \vspace{-2 mm}
      \includegraphics{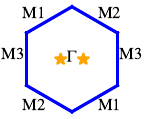}
    \end{minipage} 
    }
    & 
     \multirow[c]{4}{*}[0in]{
      \begin{minipage}[t]{0\textwidth}
      \vspace{-16 mm}
      \hspace*{-4 mm}
      \includegraphics{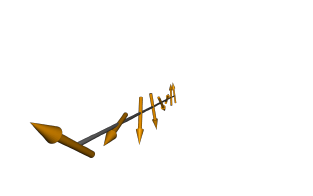}
    \end{minipage} 

    }
    \\ 
    
    &
    &
    \\
    Spiral
    &
    &
    \\ 
    
    &
    &
    \\ 
    
    &
    &
    Incommensurate coplanar order, here show an example with period 8.
    \\ 
    
    &
    &
    \\ \hline
    FP
    &
    $\bm{Q}=0 \  (\Gamma)$
    &
    \\ \hline
  \end{tabular}
  
  \caption{Hartree-Fock magnetic orders. Wave vectors $\bm{Q}_s$ of the magnetic orders are denoted as stars in the Brillouin zone. }\label{tbl:HForders}
\end{table}

\paragraph{Spin-density-wave orders.}
The numerical results we detail below show that magnetic orders are ubiquitous in the 1/2-electron-doping  phase diagram. These may be parameterized in terms of expectation values of the spin operators  $\mathbf{S}(j)=\frac{1}{2} \sum_{s,s'}c^{\dagger}_{js}\bm{\sigma}_{s,s'} c_{j s'}$, where $\bm{\sigma}=[\sigma_x, \sigma_y, \sigma_z]$ are Pauli matrices. We decompose the corresponding spin textures as
$\langle \mathbf{S}(j)\rangle = \sum_{s} \bm{m}(\bm{Q}_s)e^{i\bm{Q}_s\cdot\bm{r}_j}$. Here $\bm{r}_j$ is the coordinate of site $j$: $\bm{r}_j=r_{j,1} \bm{a}_1+ r_{j,2} \bm{a}_2$, with $\bm{a}_1$ and $\bm{a}_2$ the Bravais lattice vectors. Commensurate spin textures are periodic and have wave vectors of the form $\bm{Q}_s=s_1\bm{G}_{1}/N_{\mathrm{u},1}+s_2\bm{G}_{2}/N_{\mathrm{u},2}$, where $\bm{G}_{1}$, $\bm{G}_{2}$ are the reciprocal lattice vectors and $s_1$, $s_2$, $N_{\mathrm{u},1}$, and $N_{\mathrm{u},2}$ are integers.
In the incommensurate case, the $\bm{Q}_s$ are irrational linear combinations of the reciprocal lattice vectors, in which case there is no periodically repeating unit cell. A prototypical type of incommensurate orders is spin spirals with $\langle S_{x}(j)+iS_{y}(j)\rangle \propto e^{i\bm{Q}\cdot\bm{r}_j}$, $\langle S_{z}(j) \rangle =m_z$, where the spiral is canted if $m_z \neq 0$.

We find ground states via variational optimization of (a) Slater determinants (unrestricted HF), (b) matrix product states (DMRG), and (c) projected entangled-pair states (PEPS). These three numerical methods are complementary. We use HF for a first insight into candidate symmetry-breaking orders. By employing ODA self-consistent optimization~\cite{ODA}, we can achieve convergence in total system size ($144 \times 144$ up to $1024 \times 1024$ sites) to high precision, and deal with relatively large unit cells (up to $48 \times 48$ sites). 
In addition to our unrestricted HF study, we have also used a ``boosted frame ansatz"~\cite{IKS,suppm} to efficiently study incommensurate spiral states in HF. 
However, as it is limited to Slater determinants, HF is an uncontrolled approximation and is not guaranteed to provide reliable results, especially at strong coupling. Tensor network methods (DMRG and PEPS), on the other hand, are asymptotically exact methods, as their accuracy can be systematically improved by increasing the number of variational parameters characterized by the bond dimension $D$. DMRG is system-size limited but can reach very high accuracy. We hence use it to extrapolate to the two-dimensional (2D) thermodynamic limit from accurate small-system calculations. Variational PEPS calculations are performed directly in the 2D thermodynamic limit, with different choices of unit cell. By introducing automatic differentiation gradient descent~\cite{ADPEPS,AD2} to fermionic PEPS~\cite{fPEPS}, we are able to achieve good optimization efficiency.

\begin{figure}
\begin{center}
\includegraphics[width=0.8\columnwidth]{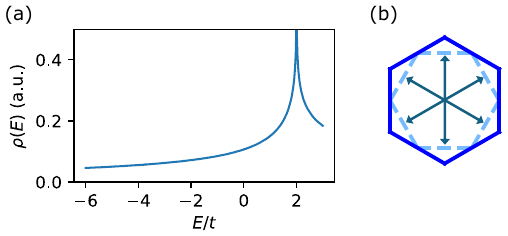}
\end{center}
\caption{(a) Density of states for the model at $U=0$. A Van Hove singularity is located at single-particle energy $E=2$. (b) Fermi surface (blue dashed line) in the Brillouin zone for  $\overline{\langle n_i\rangle}=\frac{3}{2}$. The Fermi energy is $E_F=2$. There are three nesting vectors $\mathbf{Q}_{M_i} = -\mathbf{Q}_{M_i}$ modulo reciprocal lattice vectors.}
\label{fig:band}
\end{figure}

\paragraph{1/2 electron doping: Hartree-Fock.}
At 1/2 electron doping, the noninteracting band structure has Van Hove singularities at the Fermi energy, stemming from saddle points in the dispersion at the three $M$ points (Fig.~\ref{fig:band}). Previous work~\cite{MartinBatista} has built on this to propose the emergence of a weak-coupling ``3Q'' magnetic order characterized by 
the three independent choices of wave vector $\Delta \mathbf{Q}_{ij} \equiv \mathbf{Q}_{M_i} -\mathbf{Q}_{M_j}=\epsilon_{jik}\mathbf{Q}_{M_k}$ connecting  the $M$ points $\mathbf{Q}_{M_i}, i=1,2,3$, since  
$\mathbf{Q}_{M_i} = -\mathbf{Q}_{M_i}$ modulo reciprocal lattice vectors.
The 3Q order leads to a $2\times2$ enlarged unit cell in real space. This state, that we term ``3Q-I'' for reasons that will be clear shortly, is a Chern insulator with tetrahedral spin order (Table.~\ref{tbl:HForders}) and was found in restricted HF calculations~\cite{MartinBatista}. Subsequently, it was found that within HF ferromagnetic states have a lower energy than the 3Q-I state~\cite{PasrijaKumar,ZangWang,AkagiMotome} for $U>3.5$, suggesting the breakdown of the weak-coupling picture. 

Here, we find that at intermediate interaction strengths several other magnetically ordered metallic states appear between a 3Q order and the ferromagnet. This results in a remarkably rich phase diagram, with the striking trend that the ordering vectors appear to systematically evolve from $M$ to $\Gamma$ with increasing $U$.  
The evidence for large-scale spin textures is first obtained from our unrestricted HF calculations. In our HF phase diagram [Fig.~\ref{fig:PD}(a)], we find the 3Q Chern insulator is further restricted to the range $U<2.5$. In this region, we also find a competing 3Q collinear state (Table.~\ref{tbl:HForders}) that we dub ``3Q-II'', with $\lesssim 10^{-4}$ higher energy per site.   
 This competing collinear state is distinct from that proposed in Refs.~\cite{MartinBatista, li2010metalinsulator}, as it has net spin polarization and also a charge modulation induced by the spin order, though both are an order of magnitude smaller than the spin modulation. 
Similar orders have also been found in renormalization group (RG) studies~\cite{NandkishorChern}, and the collinear state or its unpolarized sibling has been argued to be favored over the 3Q insulator at finite temperatures in both mean-field and RG studies~\cite{li2010metalinsulator,NandkishorChern}.

As $U$ is increased further, the 3Q tetrahedral order is first replaced by the ``6Q-I'' state: a noncoplanar spin texture with ordering wave vectors shifted away from the $M$ points towards  $\Gamma$. At still larger $U$, the 6Q-I order gives way to the 6Q-II order, which is collinear and has a small net magnetization and small induced charge density modulation. Our data obtained on different momentum grids indicate that the 6Q-II order is most likely commensurate and has a fixed $3 \times 3$ unit cell, and can be considered as a larger unit cell version of the competing $2 \times 2$ collinear state (the 3Q-II state) found at small $U$. 

Upon further increasing $U$, a multi-Q noncoplanar stripe (MQS) takes over from the 6Q-II state. The MQS state has an $8\times1$ unit cell and is found to have the lowest energy for a short interval of $U$. It has three pairs of ordering wave vectors $\pm \bm{Q}_i$. For even larger $U$, an incommensurate single-Q uncanted spiral phase is found to be favored until $U \sim 5.2$. Within the spiral region, the ordering wave vector varies continuously towards $\Gamma$, approaching it very closely for $U \sim 5.2$, after which the fully polarized (FP) ferromagnetic state becomes more favoured. While the ordering wave vector of the spiral is still located on the $\Gamma-M$ lines, the energy density difference by rotating to other momentum directions can be as small as $10^{-4}$. The Slater determinant of the FP state is an exact eigenstate of $H$. Since HF is variational, the region of full spin polarization is therefore bounded below by $U \sim  5.2$.

Except for the 3Q-I tetrahedral state, all other states are found to be metallic. By the Lieb-Schultz-Mattis (LSM) theorem~\cite{LIEB1961407,Oshikawa00,Hastings04}, and assuming there is no topological order, states with a fractional charge per unit cell must be metallic. For circular spiral orders, a generalized LSM theorem~\cite{IKS} further states that a fractional charge per site leads to metallic behavior. Of the states found in HF, only the 3Q-I (spin-tetrahedral), 3Q-II (collinear), and the MQS states can in principle be insulating according to the (generalized) LSM theorem. Nevertheless, we find the latter two states to be metallic. 
\begin{figure}
\begin{center}
\includegraphics[width=1\columnwidth]{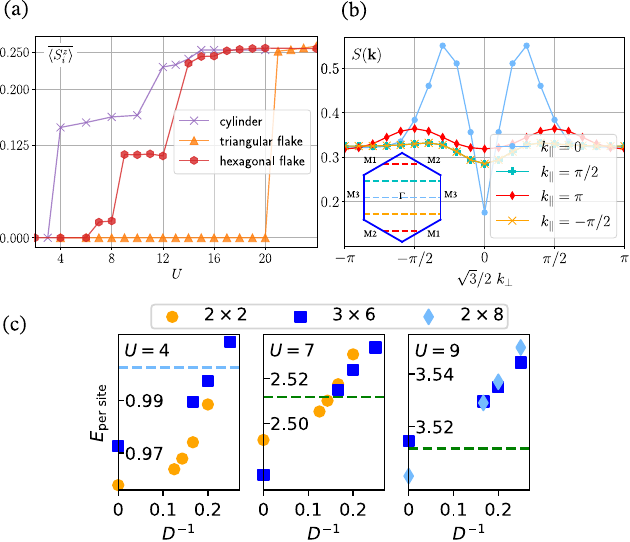}
\end{center}
\caption{Tensor network results for 1/2 electron doping.  (a,b) DMRG data. (a) Polarization $\overline {\langle S_{z}(i) \rangle}$ (polarization per site) as a function of $U$ for different geometries: four-leg zigzag cylinder with length 20, 36-site triangular flake, and 48-site hexagonal flake.  The $\overline {\langle  S_{z}(i) \rangle}$ is evaluated for the highest-weight ground state averaged without edge sites.  (b) Static structural factor $S(\bm{k})=\sum_{m}e^{i \bm{k}\cdot (\bm{r}_m-\bm{r}_n)}[\langle \bm{S}(m) \cdot \bm{S}(n)\rangle-\langle \bm{S}(m)\rangle \cdot \langle \bm{S}(n)\rangle]$ averaged for the central sites $m$ of the cylinder geometry, $U=8$. The $\bm{k}$ can be projected to two orthogonal components $k_{\perp}$ and $k_{\parallel}$. The four possible values of $k_{\parallel}$  correspond to four colored cuts in the Brillouin zone, shown in the insets. (The red dashed lines are along the same cut through M1 and M2.) (c) PEPS variational energy data for different unit cells. The dashed lines denote HF energies for comparison, with full spin polarization at $U=7,9$. Different PEPS unit cells $m \times n$  are used, which sets the maximal possible unit cells of the obtained states. The spiral-like states obtained for $2 \times 8$ PEPS unit cell at $U=9$ have smaller actual unit cells, i.e.,  $1 \times 8$. The $2 \times 2$  and  $3 \times 6$ states for $U=4,7$ are collinear. We perform extrapolation to the $D^{-1}=0$ limit based on $E(D)=E_0+ c/D^\alpha$ ansatz~\cite{suppm}.
}\label{fig:TN32}
\end{figure}

\paragraph{1/2 electron doping: DMRG.}  
We now go beyond mean-field theory and numerically test the predictions of HF with tensor network methods. The inferred schematic phase diagram obtained from both DMRG and PEPS is shown in Fig.~\ref{fig:TN32}(a). Compared to HF, the phases and their boundaries are shifted to stronger interactions, in line with the expectation for comparing  mean-field results with those of more controlled methods ~\cite{PhysRevX.3.031010,PhysRevB.103.155110}.

We use DMRG to obtain ground states on small systems (a 36-site triangular flake, a 48-site hexagonal flake, and a finite cylinder). The results indicate strong coupling ferromagnetism and demonstrate the challenges of studying the possibility of SDW orders at weaker interactions on small system sizes. Our simulations  explicitly conserve charge U$(1)$, spin SU$(2)$ ~\cite{McCulloch_2007,SciPostPhys.14.3.052}; transverse momentum~\cite{PhysRevB.93.155139} conservation is additionally used as validation for cylinder data. Broken-symmetry phases are identified by measuring the spin polarization $S_{z}$ and correlation functions. The calculated $S_{z}$ per site as a function of $U$ [Fig.~\ref{fig:TN32}(a)] is quite different for different geometries, but the critical $U$'s to reach full polarization are in reasonable agreement ($U=17 \pm 4$). This difference is not very surprising because most large-scale spin textures found in HF do not fit well on moderately sized DMRG systems. The DMRG results are therefore consistent with the existence of large-scale spin textures in the thermodynamic limit. The estimate for the critical interaction strength for a homogeneous order such as ferromagnetism, on the other hand, is probably less sensitive to finite-size effects. The finite-size limitations can also already be estimated at the HF level. In HF studies on similar small system sizes as used in DMRG,  spin textures that emerge on large systems are also absent. However, the deviation of the critical $U$ for ferromagnetism is smaller than $20\%$ relative to larger system sizes.
We note that evidence for ferromagnetism at high $U$ for similar fillings has previously also been obtained in small-system data of a simplified $t-J$ model ~\cite{YeRan}, and of a Hubbard model~\cite{DMFTTLHM,Xu2023,PhysRevB.107.235105, morera2023high}.

Our DMRG data for cylinder geometries show the existence of strong, unidirectional, and apparently-incommensurate SDW correlations in the  static structure factor [i.e., the only obvious peak is found at some ``incommensurate" $k_{\perp}$ for $k_{\parallel}=0$ in Fig.~\ref{fig:TN32}(b)]. This, combined with our finding of a nonzero $\langle S^z \rangle$ [Fig.~\ref{fig:TN32}(a)], indicates that the DMRG ground state exhibits canted spin spiral order.  
%Correspondingly, HF calculations on four-leg cylinders only find the spiral phase between the 3Q-I state at small $U$ and the FM at large $U$, even though the 3Q orders also fit on this geometry.
Correspondingly, HF calculations on four-leg cylinders only find the spiral phase at moderate $U$, even though the 3Q orders also fit on this geometry.
The DMRG results are thus again in agreement with HF, in that spiral orders are present at intermediate $U$ on the cylinder geometry, albeit in a canted form. 
%We attribute the difference between the results obtained on large tori and the cylinder to the intrinsically anisotropic nature of the latter, which we expect favors uni-directional orders such as the spin spirals. 
We attribute the difference between the results obtained on large tori and the thin cylinder to the intrinsically anisotropic nature of the latter, which does not capture the two-dimensional Fermi-surface nesting.

\paragraph{1/2 electron doping: PEPS.} In our infinite PEPS calculations,  guided by the HF results, we use the following unit cells: $2 \times 2$ (to fit  3Q), $3 \times 6$ (to fit 6Q-II and as a proxy for spirals) and $2 \times 8$ (to fit MQS and as a proxy for spirals). We consider $4\leq U \leq20$. A $2 \times 2$ state (3Q-II), large-scale SDWs (6Q-II analogue, spiral), and then FP states are found from weak to strong interactions.

By increasing $D$ up to $8$ we can systematically lower the variational energy of the PEPS states [Fig.~\ref{fig:TN32}(c)]. As a proof of improved accuracy, the $U=4$ data in Fig.~\ref{fig:TN32}(c) show that the energies of the $2\times 2$ and $3\times 6$ PEPS are lower than the HF energies. All states with a $2 \times 2$ unit cell obtained with PEPS are collinear 3Q-II states  (Table~\ref{tbl:HForders}), but with no clear net magnetization. The 3Q-I tetrahedral state with scalar chirality is not found in our calculations. However, we do not consider this state to be ruled out (see Supplemental Material~\cite{suppm}).  HF suggests that 3Q-I and 3Q-II  are very close in energy, so we take the PEPS finding of 3Q-II as evidence that one of the two 3Q states is present in the corresponding region of the phase diagram.

At $U=7$, the optimal $3 \times 6$ and $2\times 2$ unit cell PEPS are in close competition. The $3 \times 6$ PEPS has a multi-Q order, but it is difficult to establish whether it has a $3\times 6$ or $3\times 3$ collinear  unit cell, because the latter can only be strictly realized in the $D\rightarrow \infty$ limit in a U$(1)$ symmetric PEPS due to the LSM obstruction, and hence requires a careful scaling analysis with $D$.
With the $2 \times 2$ unit cell, we can reach bond dimension $D=8$, and the energy is well below the exact FP energy. This sets a better lower bound for the FP region ($U=7$) compared to HF ($U=5.2$). 

At $U=9$,  $2 \times 8$ PEPS exhibits spiral-like order with unit cells of size  $1 \times 8$. 
The spiral-like order is uncanted but has a modulated amplitude. This modulation is most likely a finite-$D$ effect
~\cite{suppm}.
The variational energies are above that of the FP state, but extrapolations in $D$ suggest that they approach the FP energy at large $D$ [see Fig.~\ref{fig:TN32}(c)]. For larger $U$, e.g., $U=12$, the only PEPS competing states are spirals and FP. The  $2 \times 2$  (not plotted) states beyond $U=9$ are nearly fully polarized. We conjecture that by increasing $U$ from 7 to 12, the effective magnetic exchange interaction between electrons has changed from antiferromagnetic to ferromagnetic coupling.  The long-wave-length spiral has nearly aligned spins for neighbors and is a compromise between ferromagnetic tendency and band dispersive energies.

\paragraph{1/2 hole doping.}The noninteracting band structure at 1/2 hole doping has an almost perfectly circular Fermi surface. Thus, weak-coupling SDWs seem unlikely and are absent in restricted HF~\cite{li2010metalinsulator,PasrijaKumar} and our unrestricted HF. Our PEPS results also indicate the absence of magnetic orders at least for $U\leq 16$ ~\cite{suppm}.

\paragraph{Discussion.} 
We now connect our results to the effective triangular-lattice systems of bilayer TMDs. Interaction strength in units of the bandwidth can be adjusted by changing the twist angle or choosing different untwisted hetero bilayers. The interaction is also long-ranged. Restricted HF results~\cite{PanWu1} indicate that with sufficiently strong long-range Coulomb interaction, insulating stripe phases can emerge for both 1/2 electron and hole doping, with ferromagnetic or antiferromagnetic order. We have performed HF including long-range interaction [$V \neq 0$ in Eq.~\eqref{eq:Hamiltonian}], and find that for the SDW region, a reasonable strength of $V$ and screening length is not sufficient to stabilize stripes or fluxed phases~\cite{Chen2024}, i.e., the existence of large-scale SDW remains unaffected by the long-range interactions. In the experiments of Refs.~\cite{Wang2020,Li2021}, no insulating phases are observed near 1/2 electron and hole doping, indicating that neither the 3Q-I spin-tetrahedral state nor stripe states are realized. The tunable metal-insulator transition at neutrality indicates that the interaction strength can be estimated as $U \sim 6-9$ (which roughly corresponds to $U \sim 2.5-3.5$ in HF), which is within the parameter regime that we have considered in this work.
For experimental investigations of SDWs, the spin-valley locking in TMDs has two important consequences: in-plane spin order will result in atomic-scale order (due to the associated inter-valley coherence), and perturbations to the TLHM break the SU$(2)$ spin rotation symmetry down to $\mathbb{Z}_2\rtimes$U$(1)$, allowing for finite-temperature long-range order.

\emph{Acknowledgements.}
We thank Steven Kivelson, Chong Wang, and Congjun Wu for discussions. Y.H. and S.A.P. acknowledge support from the European Research Council (ERC) under the European Union Horizon 2020 Research and Innovation Programme (Grant Agreement No. 804213-TMCS). M.P. is funded by the Deutsche Forschungsgemeinschaft (DFG, German Research Foundation) – Project ID 497779765. C.K. and R.R. acknowledge support by the Deutsche Forschungsgemeinschaft (DFG, German Research Foundation) under Germany's Excellence Strategy EXC-2123 QuantumFrontiers 390837967.
N.B. has received funding from the European
Research Council (ERC) under the European
Union’s Horizon 2020 Research and
Innovation Programme (Grant Agreement No.
101076597 - SIESS). P.C. acknowledges support from
the European Research Council (ERC) under the European
Union’s Horizon 2020 Research and Innovation
Programme (grant agreement No. 101001604).
 PEPS simulations are performed with the Xped library (\url{https://github.com/cpp977/Xped}). The authors would like to acknowledge the use of the University of Oxford Advanced Research Computing (ARC) facility in carrying out this work (\url{http://dx.doi.org/10.5281/zenodo.22558}). 
\bibliography{bib}

\begin{thebibliography}{81}%
\makeatletter
\providecommand \@ifxundefined [1]{%
 \@ifx{#1\undefined}
}%
\providecommand \@ifnum [1]{%
 \ifnum #1\expandafter \@firstoftwo
 \else \expandafter \@secondoftwo
 \fi
}%
\providecommand \@ifx [1]{%
 \ifx #1\expandafter \@firstoftwo
 \else \expandafter \@secondoftwo
 \fi
}%
\providecommand \natexlab [1]{#1}%
\providecommand \enquote  [1]{``#1''}%
\providecommand \bibnamefont  [1]{#1}%
\providecommand \bibfnamefont [1]{#1}%
\providecommand \citenamefont [1]{#1}%
\providecommand \href@noop [0]{\@secondoftwo}%
\providecommand \href [0]{\begingroup \@sanitize@url \@href}%
\providecommand \@href[1]{\@@startlink{#1}\@@href}%
\providecommand \@@href[1]{\endgroup#1\@@endlink}%
\providecommand \@sanitize@url [0]{\catcode `\\12\catcode `\$12\catcode `\&12\catcode `\#12\catcode `\^12\catcode `\_12\catcode `\%12\relax}%
\providecommand \@@startlink[1]{}%
\providecommand \@@endlink[0]{}%
\providecommand \url  [0]{\begingroup\@sanitize@url \@url }%
\providecommand \@url [1]{\endgroup\@href {#1}{\urlprefix }}%
\providecommand \urlprefix  [0]{URL }%
\providecommand \Eprint [0]{\href }%
\providecommand \doibase [0]{https://doi.org/}%
\providecommand \selectlanguage [0]{\@gobble}%
\providecommand \bibinfo  [0]{\@secondoftwo}%
\providecommand \bibfield  [0]{\@secondoftwo}%
\providecommand \translation [1]{[#1]}%
\providecommand \BibitemOpen [0]{}%
\providecommand \bibitemStop [0]{}%
\providecommand \bibitemNoStop [0]{.\EOS\space}%
\providecommand \EOS [0]{\spacefactor3000\relax}%
\providecommand \BibitemShut  [1]{\csname bibitem#1\endcsname}%
\let\auto@bib@innerbib\@empty
%</preamble>
\bibitem [{\citenamefont {Wannier}(1950)}]{WannierIsingtriangular}%
  \BibitemOpen
  \bibfield  {author} {\bibinfo {author} {\bibfnamefont {G.~H.}\ \bibnamefont {Wannier}},\ }\bibfield  {title} {\bibinfo {title} {{Antiferromagnetism. The Triangular Ising Net}},\ }\href {https://doi.org/10.1103/PhysRev.79.357} {\bibfield  {journal} {\bibinfo  {journal} {Phys. Rev.}\ }\textbf {\bibinfo {volume} {79}},\ \bibinfo {pages} {357} (\bibinfo {year} {1950})}\BibitemShut {NoStop}%
\bibitem [{\citenamefont {Shimizu}\ \emph {et~al.}(2003)\citenamefont {Shimizu}, \citenamefont {Miyagawa}, \citenamefont {Kanoda}, \citenamefont {Maesato},\ and\ \citenamefont {Saito}}]{Shimizu2003}%
  \BibitemOpen
  \bibfield  {author} {\bibinfo {author} {\bibfnamefont {Y.}~\bibnamefont {Shimizu}}, \bibinfo {author} {\bibfnamefont {K.}~\bibnamefont {Miyagawa}}, \bibinfo {author} {\bibfnamefont {K.}~\bibnamefont {Kanoda}}, \bibinfo {author} {\bibfnamefont {M.}~\bibnamefont {Maesato}},\ and\ \bibinfo {author} {\bibfnamefont {G.}~\bibnamefont {Saito}},\ }\bibfield  {title} {\bibinfo {title} {{Spin Liquid State in an Organic Mott Insulator with a Triangular Lattice}},\ }\href {https://doi.org/10.1103/PhysRevLett.91.107001} {\bibfield  {journal} {\bibinfo  {journal} {Phys. Rev. Lett.}\ }\textbf {\bibinfo {volume} {91}},\ \bibinfo {pages} {107001} (\bibinfo {year} {2003})}\BibitemShut {NoStop}%
\bibitem [{\citenamefont {Takada}\ \emph {et~al.}(2003)\citenamefont {Takada}, \citenamefont {Sakurai}, \citenamefont {Takayama-Muromachi}, \citenamefont {Izumi}, \citenamefont {Dilanian},\ and\ \citenamefont {Sasaki}}]{Takada2003}%
  \BibitemOpen
  \bibfield  {author} {\bibinfo {author} {\bibfnamefont {K.}~\bibnamefont {Takada}}, \bibinfo {author} {\bibfnamefont {H.}~\bibnamefont {Sakurai}}, \bibinfo {author} {\bibfnamefont {E.}~\bibnamefont {Takayama-Muromachi}}, \bibinfo {author} {\bibfnamefont {F.}~\bibnamefont {Izumi}}, \bibinfo {author} {\bibfnamefont {R.~A.}\ \bibnamefont {Dilanian}},\ and\ \bibinfo {author} {\bibfnamefont {T.}~\bibnamefont {Sasaki}},\ }\bibfield  {title} {\bibinfo {title} {{Superconductivity in two-dimensional $\mathrm{CoO}_{2}$ layers}},\ }\href {https://doi.org/10.1038/nature01450} {\bibfield  {journal} {\bibinfo  {journal} {Nature}\ }\textbf {\bibinfo {volume} {422}},\ \bibinfo {pages} {53} (\bibinfo {year} {2003})}\BibitemShut {NoStop}%
\bibitem [{\citenamefont {Hasan}\ \emph {et~al.}(2004)\citenamefont {Hasan}, \citenamefont {Chuang}, \citenamefont {Qian}, \citenamefont {Li}, \citenamefont {Kong}, \citenamefont {Kuprin}, \citenamefont {Fedorov}, \citenamefont {Kimmerling}, \citenamefont {Rotenberg}, \citenamefont {Rossnagel}, \citenamefont {Hussain}, \citenamefont {Koh}, \citenamefont {Rogado}, \citenamefont {Foo},\ and\ \citenamefont {Cava}}]{Hasan2004}%
  \BibitemOpen
  \bibfield  {author} {\bibinfo {author} {\bibfnamefont {M.~Z.}\ \bibnamefont {Hasan}}, \bibinfo {author} {\bibfnamefont {Y.-D.}\ \bibnamefont {Chuang}}, \bibinfo {author} {\bibfnamefont {D.}~\bibnamefont {Qian}}, \bibinfo {author} {\bibfnamefont {Y.~W.}\ \bibnamefont {Li}}, \bibinfo {author} {\bibfnamefont {Y.}~\bibnamefont {Kong}}, \bibinfo {author} {\bibfnamefont {A.}~\bibnamefont {Kuprin}}, \bibinfo {author} {\bibfnamefont {A.~V.}\ \bibnamefont {Fedorov}}, \bibinfo {author} {\bibfnamefont {R.}~\bibnamefont {Kimmerling}}, \bibinfo {author} {\bibfnamefont {E.}~\bibnamefont {Rotenberg}}, \bibinfo {author} {\bibfnamefont {K.}~\bibnamefont {Rossnagel}}, \bibinfo {author} {\bibfnamefont {Z.}~\bibnamefont {Hussain}}, \bibinfo {author} {\bibfnamefont {H.}~\bibnamefont {Koh}}, \bibinfo {author} {\bibfnamefont {N.~S.}\ \bibnamefont {Rogado}}, \bibinfo {author} {\bibfnamefont {M.~L.}\ \bibnamefont {Foo}},\ and\ \bibinfo {author} {\bibfnamefont {R.~J.}\ \bibnamefont {Cava}},\ }\bibfield  {title} {\bibinfo {title}
  {{Fermi Surface and Quasiparticle Dynamics of ${\mathrm{Na}}_{0.7}{\mathrm{CoO}}_{2}$ Investigated by Angle-Resolved Photoemission Spectroscopy}},\ }\href {https://doi.org/10.1103/PhysRevLett.92.246402} {\bibfield  {journal} {\bibinfo  {journal} {Phys. Rev. Lett.}\ }\textbf {\bibinfo {volume} {92}},\ \bibinfo {pages} {246402} (\bibinfo {year} {2004})}\BibitemShut {NoStop}%
\bibitem [{\citenamefont {Foo}\ \emph {et~al.}(2004)\citenamefont {Foo}, \citenamefont {Wang}, \citenamefont {Watauchi}, \citenamefont {Zandbergen}, \citenamefont {He}, \citenamefont {Cava},\ and\ \citenamefont {Ong}}]{Foo2004}%
  \BibitemOpen
  \bibfield  {author} {\bibinfo {author} {\bibfnamefont {M.~L.}\ \bibnamefont {Foo}}, \bibinfo {author} {\bibfnamefont {Y.}~\bibnamefont {Wang}}, \bibinfo {author} {\bibfnamefont {S.}~\bibnamefont {Watauchi}}, \bibinfo {author} {\bibfnamefont {H.~W.}\ \bibnamefont {Zandbergen}}, \bibinfo {author} {\bibfnamefont {T.}~\bibnamefont {He}}, \bibinfo {author} {\bibfnamefont {R.~J.}\ \bibnamefont {Cava}},\ and\ \bibinfo {author} {\bibfnamefont {N.~P.}\ \bibnamefont {Ong}},\ }\bibfield  {title} {\bibinfo {title} {{Charge Ordering, Commensurability, and Metallicity in the Phase Diagram of the Layered ${\mathrm{Na}}_{x}\mathrm{Co}{\mathrm{O}}_{2}$}},\ }\href {https://doi.org/10.1103/PhysRevLett.92.247001} {\bibfield  {journal} {\bibinfo  {journal} {Phys. Rev. Lett.}\ }\textbf {\bibinfo {volume} {92}},\ \bibinfo {pages} {247001} (\bibinfo {year} {2004})}\BibitemShut {NoStop}%
\bibitem [{\citenamefont {Ga\ifmmode \check{s}\else \v{s}\fi{}parovi\ifmmode~\acute{c}\else \'{c}\fi{}}\ \emph {et~al.}(2006)\citenamefont {Ga\ifmmode \check{s}\else \v{s}\fi{}parovi\ifmmode~\acute{c}\else \'{c}\fi{}}, \citenamefont {Ott}, \citenamefont {Cho}, \citenamefont {Chou}, \citenamefont {Chu}, \citenamefont {Lynn},\ and\ \citenamefont {Lee}}]{Lee06}%
  \BibitemOpen
  \bibfield  {author} {\bibinfo {author} {\bibfnamefont {G.}~\bibnamefont {Ga\ifmmode \check{s}\else \v{s}\fi{}parovi\ifmmode~\acute{c}\else \'{c}\fi{}}}, \bibinfo {author} {\bibfnamefont {R.~A.}\ \bibnamefont {Ott}}, \bibinfo {author} {\bibfnamefont {J.-H.}\ \bibnamefont {Cho}}, \bibinfo {author} {\bibfnamefont {F.~C.}\ \bibnamefont {Chou}}, \bibinfo {author} {\bibfnamefont {Y.}~\bibnamefont {Chu}}, \bibinfo {author} {\bibfnamefont {J.~W.}\ \bibnamefont {Lynn}},\ and\ \bibinfo {author} {\bibfnamefont {Y.~S.}\ \bibnamefont {Lee}},\ }\bibfield  {title} {\bibinfo {title} {{Neutron Scattering Study of Novel Magnetic Order in ${\mathrm{Na}}_{0.5}{\mathrm{CoO}}_{2}$}},\ }\href {https://doi.org/10.1103/PhysRevLett.96.046403} {\bibfield  {journal} {\bibinfo  {journal} {Phys. Rev. Lett.}\ }\textbf {\bibinfo {volume} {96}},\ \bibinfo {pages} {046403} (\bibinfo {year} {2006})}\BibitemShut {NoStop}%
\bibitem [{\citenamefont {Zhou}\ \emph {et~al.}(2011)\citenamefont {Zhou}, \citenamefont {Choi}, \citenamefont {Li}, \citenamefont {Balicas}, \citenamefont {Wiebe}, \citenamefont {Qiu}, \citenamefont {Copley},\ and\ \citenamefont {Gardner}}]{Zhou11}%
  \BibitemOpen
  \bibfield  {author} {\bibinfo {author} {\bibfnamefont {H.~D.}\ \bibnamefont {Zhou}}, \bibinfo {author} {\bibfnamefont {E.~S.}\ \bibnamefont {Choi}}, \bibinfo {author} {\bibfnamefont {G.}~\bibnamefont {Li}}, \bibinfo {author} {\bibfnamefont {L.}~\bibnamefont {Balicas}}, \bibinfo {author} {\bibfnamefont {C.~R.}\ \bibnamefont {Wiebe}}, \bibinfo {author} {\bibfnamefont {Y.}~\bibnamefont {Qiu}}, \bibinfo {author} {\bibfnamefont {J.~R.~D.}\ \bibnamefont {Copley}},\ and\ \bibinfo {author} {\bibfnamefont {J.~S.}\ \bibnamefont {Gardner}},\ }\bibfield  {title} {\bibinfo {title} {{Spin Liquid State in the $S=1/2$ Triangular Lattice ${\mathrm{Ba}}_{3}{\mathrm{CuSb}}_{2}{\mathrm{O}}_{9}$}},\ }\href {https://doi.org/10.1103/PhysRevLett.106.147204} {\bibfield  {journal} {\bibinfo  {journal} {Phys. Rev. Lett.}\ }\textbf {\bibinfo {volume} {106}},\ \bibinfo {pages} {147204} (\bibinfo {year} {2011})}\BibitemShut {NoStop}%
\bibitem [{\citenamefont {Zvyagin}\ \emph {et~al.}(2019)\citenamefont {Zvyagin}, \citenamefont {Graf}, \citenamefont {Sakurai}, \citenamefont {Kimura}, \citenamefont {Nojiri}, \citenamefont {Wosnitza}, \citenamefont {Ohta}, \citenamefont {Ono},\ and\ \citenamefont {Tanaka}}]{Zvyagin2019}%
  \BibitemOpen
  \bibfield  {author} {\bibinfo {author} {\bibfnamefont {S.~A.}\ \bibnamefont {Zvyagin}}, \bibinfo {author} {\bibfnamefont {D.}~\bibnamefont {Graf}}, \bibinfo {author} {\bibfnamefont {T.}~\bibnamefont {Sakurai}}, \bibinfo {author} {\bibfnamefont {S.}~\bibnamefont {Kimura}}, \bibinfo {author} {\bibfnamefont {H.}~\bibnamefont {Nojiri}}, \bibinfo {author} {\bibfnamefont {J.}~\bibnamefont {Wosnitza}}, \bibinfo {author} {\bibfnamefont {H.}~\bibnamefont {Ohta}}, \bibinfo {author} {\bibfnamefont {T.}~\bibnamefont {Ono}},\ and\ \bibinfo {author} {\bibfnamefont {H.}~\bibnamefont {Tanaka}},\ }\bibfield  {title} {\bibinfo {title} {{Pressure-tuning the quantum spin Hamiltonian of the triangular lattice antiferromagnet $\mathrm{Cs}\mathrm{CuCl}_4$}},\ }\href {https://doi.org/10.1038/s41467-019-09071-7} {\bibfield  {journal} {\bibinfo  {journal} {Nature Communications}\ }\textbf {\bibinfo {volume} {10}},\ \bibinfo {pages} {1064} (\bibinfo {year} {2019})}\BibitemShut {NoStop}%
\bibitem [{\citenamefont {Wu}\ \emph {et~al.}(2018)\citenamefont {Wu}, \citenamefont {Lovorn}, \citenamefont {Tutuc},\ and\ \citenamefont {MacDonald}}]{Wu}%
  \BibitemOpen
  \bibfield  {author} {\bibinfo {author} {\bibfnamefont {F.}~\bibnamefont {Wu}}, \bibinfo {author} {\bibfnamefont {T.}~\bibnamefont {Lovorn}}, \bibinfo {author} {\bibfnamefont {E.}~\bibnamefont {Tutuc}},\ and\ \bibinfo {author} {\bibfnamefont {A.~H.}\ \bibnamefont {MacDonald}},\ }\bibfield  {title} {\bibinfo {title} {{Hubbard Model Physics in Transition Metal Dichalcogenide Moir\'e Bands}},\ }\href {https://doi.org/10.1103/PhysRevLett.121.026402} {\bibfield  {journal} {\bibinfo  {journal} {Phys. Rev. Lett.}\ }\textbf {\bibinfo {volume} {121}},\ \bibinfo {pages} {026402} (\bibinfo {year} {2018})}\BibitemShut {NoStop}%
\bibitem [{\citenamefont {Zhang}\ \emph {et~al.}(2017)\citenamefont {Zhang}, \citenamefont {Chuu}, \citenamefont {Ren}, \citenamefont {Li}, \citenamefont {Li}, \citenamefont {Jin}, \citenamefont {Chou},\ and\ \citenamefont {Shih}}]{Zhang2017}%
  \BibitemOpen
  \bibfield  {author} {\bibinfo {author} {\bibfnamefont {C.}~\bibnamefont {Zhang}}, \bibinfo {author} {\bibfnamefont {C.-P.}\ \bibnamefont {Chuu}}, \bibinfo {author} {\bibfnamefont {X.}~\bibnamefont {Ren}}, \bibinfo {author} {\bibfnamefont {M.-Y.}\ \bibnamefont {Li}}, \bibinfo {author} {\bibfnamefont {L.-J.}\ \bibnamefont {Li}}, \bibinfo {author} {\bibfnamefont {C.}~\bibnamefont {Jin}}, \bibinfo {author} {\bibfnamefont {M.-Y.}\ \bibnamefont {Chou}},\ and\ \bibinfo {author} {\bibfnamefont {C.-K.}\ \bibnamefont {Shih}},\ }\bibfield  {title} {\bibinfo {title} {{Interlayer couplings, Moir\'e patterns, and 2D electronic superlattices in $\mathrm{MoS}_2/\mathrm{WSe}_2$ hetero-bilayers}},\ }\href {https://doi.org/10.1126/sciadv.1601459} {\bibfield  {journal} {\bibinfo  {journal} {Science Advances}\ }\textbf {\bibinfo {volume} {3}},\ \bibinfo {pages} {e1601459} (\bibinfo {year} {2017})}\BibitemShut {NoStop}%
\bibitem [{\citenamefont {Pan}\ \emph {et~al.}(2018)\citenamefont {Pan}, \citenamefont {F{\"o}lsch}, \citenamefont {Nie}, \citenamefont {Waters}, \citenamefont {Lin}, \citenamefont {Jariwala}, \citenamefont {Zhang}, \citenamefont {Cho}, \citenamefont {Robinson},\ and\ \citenamefont {Feenstra}}]{Pan2018}%
  \BibitemOpen
  \bibfield  {author} {\bibinfo {author} {\bibfnamefont {Y.}~\bibnamefont {Pan}}, \bibinfo {author} {\bibfnamefont {S.}~\bibnamefont {F{\"o}lsch}}, \bibinfo {author} {\bibfnamefont {Y.}~\bibnamefont {Nie}}, \bibinfo {author} {\bibfnamefont {D.}~\bibnamefont {Waters}}, \bibinfo {author} {\bibfnamefont {Y.-C.}\ \bibnamefont {Lin}}, \bibinfo {author} {\bibfnamefont {B.}~\bibnamefont {Jariwala}}, \bibinfo {author} {\bibfnamefont {K.}~\bibnamefont {Zhang}}, \bibinfo {author} {\bibfnamefont {K.}~\bibnamefont {Cho}}, \bibinfo {author} {\bibfnamefont {J.~A.}\ \bibnamefont {Robinson}},\ and\ \bibinfo {author} {\bibfnamefont {R.~M.}\ \bibnamefont {Feenstra}},\ }\bibfield  {title} {\bibinfo {title} {{Quantum-Confined Electronic States Arising from the Moir{\'e} Pattern of $\mathrm{MoS}_{2}\mathrm{--WSe}_{2}$ Heterobilayers}},\ }\href {https://doi.org/10.1021/acs.nanolett.7b05125} {\bibfield  {journal} {\bibinfo  {journal} {Nano Letters}\ }\textbf {\bibinfo {volume} {18}},\ \bibinfo {pages} {1849} (\bibinfo {year}
  {2018})}\BibitemShut {NoStop}%
\bibitem [{\citenamefont {Wang}\ \emph {et~al.}(2020)\citenamefont {Wang}, \citenamefont {Shih}, \citenamefont {Ghiotto}, \citenamefont {Xian}, \citenamefont {Rhodes}, \citenamefont {Tan}, \citenamefont {Claassen}, \citenamefont {Kennes}, \citenamefont {Bai}, \citenamefont {Kim}, \citenamefont {Watanabe}, \citenamefont {Taniguchi}, \citenamefont {Zhu}, \citenamefont {Hone}, \citenamefont {Rubio}, \citenamefont {Pasupathy},\ and\ \citenamefont {Dean}}]{Wang2020}%
  \BibitemOpen
  \bibfield  {author} {\bibinfo {author} {\bibfnamefont {L.}~\bibnamefont {Wang}}, \bibinfo {author} {\bibfnamefont {E.-M.}\ \bibnamefont {Shih}}, \bibinfo {author} {\bibfnamefont {A.}~\bibnamefont {Ghiotto}}, \bibinfo {author} {\bibfnamefont {L.}~\bibnamefont {Xian}}, \bibinfo {author} {\bibfnamefont {D.~A.}\ \bibnamefont {Rhodes}}, \bibinfo {author} {\bibfnamefont {C.}~\bibnamefont {Tan}}, \bibinfo {author} {\bibfnamefont {M.}~\bibnamefont {Claassen}}, \bibinfo {author} {\bibfnamefont {D.~M.}\ \bibnamefont {Kennes}}, \bibinfo {author} {\bibfnamefont {Y.}~\bibnamefont {Bai}}, \bibinfo {author} {\bibfnamefont {B.}~\bibnamefont {Kim}}, \bibinfo {author} {\bibfnamefont {K.}~\bibnamefont {Watanabe}}, \bibinfo {author} {\bibfnamefont {T.}~\bibnamefont {Taniguchi}}, \bibinfo {author} {\bibfnamefont {X.}~\bibnamefont {Zhu}}, \bibinfo {author} {\bibfnamefont {J.}~\bibnamefont {Hone}}, \bibinfo {author} {\bibfnamefont {A.}~\bibnamefont {Rubio}}, \bibinfo {author} {\bibfnamefont {A.~N.}\ \bibnamefont {Pasupathy}},\
  and\ \bibinfo {author} {\bibfnamefont {C.~R.}\ \bibnamefont {Dean}},\ }\bibfield  {title} {\bibinfo {title} {{Correlated electronic phases in twisted bilayer transition metal dichalcogenides}},\ }\href {https://doi.org/10.1038/s41563-020-0708-6} {\bibfield  {journal} {\bibinfo  {journal} {Nature Materials}\ }\textbf {\bibinfo {volume} {19}},\ \bibinfo {pages} {861} (\bibinfo {year} {2020})}\BibitemShut {NoStop}%
\bibitem [{\citenamefont {Regan}\ \emph {et~al.}(2020)\citenamefont {Regan}, \citenamefont {Wang}, \citenamefont {Jin}, \citenamefont {Bakti~Utama}, \citenamefont {Gao}, \citenamefont {Wei}, \citenamefont {Zhao}, \citenamefont {Zhao}, \citenamefont {Zhang}, \citenamefont {Yumigeta}, \citenamefont {Blei}, \citenamefont {Carlstr{\"o}m}, \citenamefont {Watanabe}, \citenamefont {Taniguchi}, \citenamefont {Tongay}, \citenamefont {Crommie}, \citenamefont {Zettl},\ and\ \citenamefont {Wang}}]{Regan2020}%
  \BibitemOpen
  \bibfield  {author} {\bibinfo {author} {\bibfnamefont {E.~C.}\ \bibnamefont {Regan}}, \bibinfo {author} {\bibfnamefont {D.}~\bibnamefont {Wang}}, \bibinfo {author} {\bibfnamefont {C.}~\bibnamefont {Jin}}, \bibinfo {author} {\bibfnamefont {M.~I.}\ \bibnamefont {Bakti~Utama}}, \bibinfo {author} {\bibfnamefont {B.}~\bibnamefont {Gao}}, \bibinfo {author} {\bibfnamefont {X.}~\bibnamefont {Wei}}, \bibinfo {author} {\bibfnamefont {S.}~\bibnamefont {Zhao}}, \bibinfo {author} {\bibfnamefont {W.}~\bibnamefont {Zhao}}, \bibinfo {author} {\bibfnamefont {Z.}~\bibnamefont {Zhang}}, \bibinfo {author} {\bibfnamefont {K.}~\bibnamefont {Yumigeta}}, \bibinfo {author} {\bibfnamefont {M.}~\bibnamefont {Blei}}, \bibinfo {author} {\bibfnamefont {J.~D.}\ \bibnamefont {Carlstr{\"o}m}}, \bibinfo {author} {\bibfnamefont {K.}~\bibnamefont {Watanabe}}, \bibinfo {author} {\bibfnamefont {T.}~\bibnamefont {Taniguchi}}, \bibinfo {author} {\bibfnamefont {S.}~\bibnamefont {Tongay}}, \bibinfo {author} {\bibfnamefont {M.}~\bibnamefont
  {Crommie}}, \bibinfo {author} {\bibfnamefont {A.}~\bibnamefont {Zettl}},\ and\ \bibinfo {author} {\bibfnamefont {F.}~\bibnamefont {Wang}},\ }\bibfield  {title} {\bibinfo {title} {{Mott and generalized Wigner crystal states in $\mathrm{WSe}_{2}\mathrm{/WS}_{2}$ moir{\'e} superlattices}},\ }\href {https://doi.org/10.1038/s41586-020-2092-4} {\bibfield  {journal} {\bibinfo  {journal} {Nature}\ }\textbf {\bibinfo {volume} {579}},\ \bibinfo {pages} {359} (\bibinfo {year} {2020})}\BibitemShut {NoStop}%
\bibitem [{\citenamefont {Tang}\ \emph {et~al.}(2020)\citenamefont {Tang}, \citenamefont {Li}, \citenamefont {Li}, \citenamefont {Xu}, \citenamefont {Liu}, \citenamefont {Barmak}, \citenamefont {Watanabe}, \citenamefont {Taniguchi}, \citenamefont {MacDonald}, \citenamefont {Shan},\ and\ \citenamefont {Mak}}]{Tang2020}%
  \BibitemOpen
  \bibfield  {author} {\bibinfo {author} {\bibfnamefont {Y.}~\bibnamefont {Tang}}, \bibinfo {author} {\bibfnamefont {L.}~\bibnamefont {Li}}, \bibinfo {author} {\bibfnamefont {T.}~\bibnamefont {Li}}, \bibinfo {author} {\bibfnamefont {Y.}~\bibnamefont {Xu}}, \bibinfo {author} {\bibfnamefont {S.}~\bibnamefont {Liu}}, \bibinfo {author} {\bibfnamefont {K.}~\bibnamefont {Barmak}}, \bibinfo {author} {\bibfnamefont {K.}~\bibnamefont {Watanabe}}, \bibinfo {author} {\bibfnamefont {T.}~\bibnamefont {Taniguchi}}, \bibinfo {author} {\bibfnamefont {A.~H.}\ \bibnamefont {MacDonald}}, \bibinfo {author} {\bibfnamefont {J.}~\bibnamefont {Shan}},\ and\ \bibinfo {author} {\bibfnamefont {K.~F.}\ \bibnamefont {Mak}},\ }\bibfield  {title} {\bibinfo {title} {{Simulation of Hubbard model physics in $\mathrm{WSe}_{2}\mathrm{/WS}_{2}$ moir{\'e} superlattices}},\ }\href {https://doi.org/10.1038/s41586-020-2085-3} {\bibfield  {journal} {\bibinfo  {journal} {Nature}\ }\textbf {\bibinfo {volume} {579}},\ \bibinfo {pages} {353} (\bibinfo
  {year} {2020})}\BibitemShut {NoStop}%
\bibitem [{\citenamefont {Huang}\ \emph {et~al.}(2021)\citenamefont {Huang}, \citenamefont {Wang}, \citenamefont {Miao}, \citenamefont {Wang}, \citenamefont {Li}, \citenamefont {Lian}, \citenamefont {Taniguchi}, \citenamefont {Watanabe}, \citenamefont {Okamoto}, \citenamefont {Xiao}, \citenamefont {Shi},\ and\ \citenamefont {Cui}}]{Huang2021}%
  \BibitemOpen
  \bibfield  {author} {\bibinfo {author} {\bibfnamefont {X.}~\bibnamefont {Huang}}, \bibinfo {author} {\bibfnamefont {T.}~\bibnamefont {Wang}}, \bibinfo {author} {\bibfnamefont {S.}~\bibnamefont {Miao}}, \bibinfo {author} {\bibfnamefont {C.}~\bibnamefont {Wang}}, \bibinfo {author} {\bibfnamefont {Z.}~\bibnamefont {Li}}, \bibinfo {author} {\bibfnamefont {Z.}~\bibnamefont {Lian}}, \bibinfo {author} {\bibfnamefont {T.}~\bibnamefont {Taniguchi}}, \bibinfo {author} {\bibfnamefont {K.}~\bibnamefont {Watanabe}}, \bibinfo {author} {\bibfnamefont {S.}~\bibnamefont {Okamoto}}, \bibinfo {author} {\bibfnamefont {D.}~\bibnamefont {Xiao}}, \bibinfo {author} {\bibfnamefont {S.-F.}\ \bibnamefont {Shi}},\ and\ \bibinfo {author} {\bibfnamefont {Y.-T.}\ \bibnamefont {Cui}},\ }\bibfield  {title} {\bibinfo {title} {{Correlated insulating states at fractional fillings of the $\mathrm{WS}_{2}\mathrm{/WSe}_{2}$ moir{\'e} lattice}},\ }\href {https://doi.org/10.1038/s41567-021-01171-w} {\bibfield  {journal} {\bibinfo  {journal}
  {Nature Physics}\ }\textbf {\bibinfo {volume} {17}},\ \bibinfo {pages} {715} (\bibinfo {year} {2021})}\BibitemShut {NoStop}%
\bibitem [{\citenamefont {Jin}\ \emph {et~al.}(2021)\citenamefont {Jin}, \citenamefont {Tao}, \citenamefont {Li}, \citenamefont {Xu}, \citenamefont {Tang}, \citenamefont {Zhu}, \citenamefont {Liu}, \citenamefont {Watanabe}, \citenamefont {Taniguchi}, \citenamefont {Hone}, \citenamefont {Fu}, \citenamefont {Shan},\ and\ \citenamefont {Mak}}]{Jin2021}%
  \BibitemOpen
  \bibfield  {author} {\bibinfo {author} {\bibfnamefont {C.}~\bibnamefont {Jin}}, \bibinfo {author} {\bibfnamefont {Z.}~\bibnamefont {Tao}}, \bibinfo {author} {\bibfnamefont {T.}~\bibnamefont {Li}}, \bibinfo {author} {\bibfnamefont {Y.}~\bibnamefont {Xu}}, \bibinfo {author} {\bibfnamefont {Y.}~\bibnamefont {Tang}}, \bibinfo {author} {\bibfnamefont {J.}~\bibnamefont {Zhu}}, \bibinfo {author} {\bibfnamefont {S.}~\bibnamefont {Liu}}, \bibinfo {author} {\bibfnamefont {K.}~\bibnamefont {Watanabe}}, \bibinfo {author} {\bibfnamefont {T.}~\bibnamefont {Taniguchi}}, \bibinfo {author} {\bibfnamefont {J.~C.}\ \bibnamefont {Hone}}, \bibinfo {author} {\bibfnamefont {L.}~\bibnamefont {Fu}}, \bibinfo {author} {\bibfnamefont {J.}~\bibnamefont {Shan}},\ and\ \bibinfo {author} {\bibfnamefont {K.~F.}\ \bibnamefont {Mak}},\ }\bibfield  {title} {\bibinfo {title} {{Stripe phases in $\mathrm{WSe}_{2}\mathrm{/WS}_{2}$ moir{\'e} superlattices}},\ }\href {https://doi.org/10.1038/s41563-021-00959-8} {\bibfield  {journal} {\bibinfo
  {journal} {Nature Materials}\ }\textbf {\bibinfo {volume} {20}},\ \bibinfo {pages} {940} (\bibinfo {year} {2021})}\BibitemShut {NoStop}%
\bibitem [{\citenamefont {Li}\ \emph {et~al.}(2021)\citenamefont {Li}, \citenamefont {Jiang}, \citenamefont {Li}, \citenamefont {Zhang}, \citenamefont {Kang}, \citenamefont {Zhu}, \citenamefont {Watanabe}, \citenamefont {Taniguchi}, \citenamefont {Chowdhury}, \citenamefont {Fu}, \citenamefont {Shan},\ and\ \citenamefont {Mak}}]{Li2021}%
  \BibitemOpen
  \bibfield  {author} {\bibinfo {author} {\bibfnamefont {T.}~\bibnamefont {Li}}, \bibinfo {author} {\bibfnamefont {S.}~\bibnamefont {Jiang}}, \bibinfo {author} {\bibfnamefont {L.}~\bibnamefont {Li}}, \bibinfo {author} {\bibfnamefont {Y.}~\bibnamefont {Zhang}}, \bibinfo {author} {\bibfnamefont {K.}~\bibnamefont {Kang}}, \bibinfo {author} {\bibfnamefont {J.}~\bibnamefont {Zhu}}, \bibinfo {author} {\bibfnamefont {K.}~\bibnamefont {Watanabe}}, \bibinfo {author} {\bibfnamefont {T.}~\bibnamefont {Taniguchi}}, \bibinfo {author} {\bibfnamefont {D.}~\bibnamefont {Chowdhury}}, \bibinfo {author} {\bibfnamefont {L.}~\bibnamefont {Fu}}, \bibinfo {author} {\bibfnamefont {J.}~\bibnamefont {Shan}},\ and\ \bibinfo {author} {\bibfnamefont {K.~F.}\ \bibnamefont {Mak}},\ }\bibfield  {title} {\bibinfo {title} {{Continuous Mott transition in semiconductor moir{\'e} superlattices}},\ }\href {https://doi.org/10.1038/s41586-021-03853-0} {\bibfield  {journal} {\bibinfo  {journal} {Nature}\ }\textbf {\bibinfo {volume} {597}},\ \bibinfo
  {pages} {350} (\bibinfo {year} {2021})}\BibitemShut {NoStop}%
\bibitem [{\citenamefont {Xu}\ \emph {et~al.}(2022)\citenamefont {Xu}, \citenamefont {Kang}, \citenamefont {Watanabe}, \citenamefont {Taniguchi}, \citenamefont {Mak},\ and\ \citenamefont {Shan}}]{Xu2022}%
  \BibitemOpen
  \bibfield  {author} {\bibinfo {author} {\bibfnamefont {Y.}~\bibnamefont {Xu}}, \bibinfo {author} {\bibfnamefont {K.}~\bibnamefont {Kang}}, \bibinfo {author} {\bibfnamefont {K.}~\bibnamefont {Watanabe}}, \bibinfo {author} {\bibfnamefont {T.}~\bibnamefont {Taniguchi}}, \bibinfo {author} {\bibfnamefont {K.~F.}\ \bibnamefont {Mak}},\ and\ \bibinfo {author} {\bibfnamefont {J.}~\bibnamefont {Shan}},\ }\bibfield  {title} {\bibinfo {title} {{A tunable bilayer Hubbard model in twisted $\mathrm{WSe}_2$}},\ }\href {https://doi.org/10.1038/s41565-022-01180-7} {\bibfield  {journal} {\bibinfo  {journal} {Nature Nanotechnology}\ }\textbf {\bibinfo {volume} {17}},\ \bibinfo {pages} {934} (\bibinfo {year} {2022})}\BibitemShut {NoStop}%
\bibitem [{\citenamefont {Tang}\ \emph {et~al.}(2023)\citenamefont {Tang}, \citenamefont {Su}, \citenamefont {Li}, \citenamefont {Xu}, \citenamefont {Liu}, \citenamefont {Watanabe}, \citenamefont {Taniguchi}, \citenamefont {Hone}, \citenamefont {Jian}, \citenamefont {Xu}, \citenamefont {Mak},\ and\ \citenamefont {Shan}}]{Tang2023}%
  \BibitemOpen
  \bibfield  {author} {\bibinfo {author} {\bibfnamefont {Y.}~\bibnamefont {Tang}}, \bibinfo {author} {\bibfnamefont {K.}~\bibnamefont {Su}}, \bibinfo {author} {\bibfnamefont {L.}~\bibnamefont {Li}}, \bibinfo {author} {\bibfnamefont {Y.}~\bibnamefont {Xu}}, \bibinfo {author} {\bibfnamefont {S.}~\bibnamefont {Liu}}, \bibinfo {author} {\bibfnamefont {K.}~\bibnamefont {Watanabe}}, \bibinfo {author} {\bibfnamefont {T.}~\bibnamefont {Taniguchi}}, \bibinfo {author} {\bibfnamefont {J.}~\bibnamefont {Hone}}, \bibinfo {author} {\bibfnamefont {C.-M.}\ \bibnamefont {Jian}}, \bibinfo {author} {\bibfnamefont {C.}~\bibnamefont {Xu}}, \bibinfo {author} {\bibfnamefont {K.~F.}\ \bibnamefont {Mak}},\ and\ \bibinfo {author} {\bibfnamefont {J.}~\bibnamefont {Shan}},\ }\bibfield  {title} {\bibinfo {title} {{Evidence of frustrated magnetic interactions in a Wigner--Mott insulator}},\ }\bibfield  {journal} {\bibinfo  {journal} {Nature Nanotechnology}\ }\href {https://doi.org/10.1038/s41565-022-01309-8} {10.1038/s41565-022-01309-8}
  (\bibinfo {year} {2023})\BibitemShut {NoStop}%
\bibitem [{\citenamefont {Ciorciaro}\ \emph {et~al.}(2023)\citenamefont {Ciorciaro}, \citenamefont {Smole{\'{n}}ski}, \citenamefont {Morera}, \citenamefont {Kiper}, \citenamefont {Hiestand}, \citenamefont {Kroner}, \citenamefont {Zhang}, \citenamefont {Watanabe}, \citenamefont {Taniguchi}, \citenamefont {Demler},\ and\ \citenamefont {{\.{I}}mamo{\u{g}}lu}}]{ciorciaro2023kinetic}%
  \BibitemOpen
  \bibfield  {author} {\bibinfo {author} {\bibfnamefont {L.}~\bibnamefont {Ciorciaro}}, \bibinfo {author} {\bibfnamefont {T.}~\bibnamefont {Smole{\'{n}}ski}}, \bibinfo {author} {\bibfnamefont {I.}~\bibnamefont {Morera}}, \bibinfo {author} {\bibfnamefont {N.}~\bibnamefont {Kiper}}, \bibinfo {author} {\bibfnamefont {S.}~\bibnamefont {Hiestand}}, \bibinfo {author} {\bibfnamefont {M.}~\bibnamefont {Kroner}}, \bibinfo {author} {\bibfnamefont {Y.}~\bibnamefont {Zhang}}, \bibinfo {author} {\bibfnamefont {K.}~\bibnamefont {Watanabe}}, \bibinfo {author} {\bibfnamefont {T.}~\bibnamefont {Taniguchi}}, \bibinfo {author} {\bibfnamefont {E.}~\bibnamefont {Demler}},\ and\ \bibinfo {author} {\bibfnamefont {A.}~\bibnamefont {{\.{I}}mamo{\u{g}}lu}},\ }\bibfield  {title} {\bibinfo {title} {{Kinetic magnetism in triangular moir{\'e} materials}},\ }\href {https://doi.org/10.1038/s41586-023-06633-0} {\bibfield  {journal} {\bibinfo  {journal} {Nature}\ }\textbf {\bibinfo {volume} {623}},\ \bibinfo {pages} {509} (\bibinfo {year}
  {2023})}\BibitemShut {NoStop}%
\bibitem [{\citenamefont {Xu}\ \emph {et~al.}(2023)\citenamefont {Xu}, \citenamefont {Kendrick}, \citenamefont {Kale}, \citenamefont {Gang}, \citenamefont {Ji}, \citenamefont {Scalettar}, \citenamefont {Lebrat},\ and\ \citenamefont {Greiner}}]{Xu2023}%
  \BibitemOpen
  \bibfield  {author} {\bibinfo {author} {\bibfnamefont {M.}~\bibnamefont {Xu}}, \bibinfo {author} {\bibfnamefont {L.~H.}\ \bibnamefont {Kendrick}}, \bibinfo {author} {\bibfnamefont {A.}~\bibnamefont {Kale}}, \bibinfo {author} {\bibfnamefont {Y.}~\bibnamefont {Gang}}, \bibinfo {author} {\bibfnamefont {G.}~\bibnamefont {Ji}}, \bibinfo {author} {\bibfnamefont {R.~T.}\ \bibnamefont {Scalettar}}, \bibinfo {author} {\bibfnamefont {M.}~\bibnamefont {Lebrat}},\ and\ \bibinfo {author} {\bibfnamefont {M.}~\bibnamefont {Greiner}},\ }\bibfield  {title} {\bibinfo {title} {{Frustration- and doping-induced magnetism in a Fermi--Hubbard simulator}},\ }\href {https://doi.org/10.1038/s41586-023-06280-5} {\bibfield  {journal} {\bibinfo  {journal} {Nature}\ }\textbf {\bibinfo {volume} {620}},\ \bibinfo {pages} {971} (\bibinfo {year} {2023})}\BibitemShut {NoStop}%
\bibitem [{\citenamefont {Lebrat}\ \emph {et~al.}(2024)\citenamefont {Lebrat}, \citenamefont {Xu}, \citenamefont {Kendrick}, \citenamefont {Kale}, \citenamefont {Gang}, \citenamefont {Seetharaman}, \citenamefont {Morera}, \citenamefont {Khatami}, \citenamefont {Demler},\ and\ \citenamefont {Greiner}}]{lebrat2023observation}%
  \BibitemOpen
  \bibfield  {author} {\bibinfo {author} {\bibfnamefont {M.}~\bibnamefont {Lebrat}}, \bibinfo {author} {\bibfnamefont {M.}~\bibnamefont {Xu}}, \bibinfo {author} {\bibfnamefont {L.~H.}\ \bibnamefont {Kendrick}}, \bibinfo {author} {\bibfnamefont {A.}~\bibnamefont {Kale}}, \bibinfo {author} {\bibfnamefont {Y.}~\bibnamefont {Gang}}, \bibinfo {author} {\bibfnamefont {P.}~\bibnamefont {Seetharaman}}, \bibinfo {author} {\bibfnamefont {I.}~\bibnamefont {Morera}}, \bibinfo {author} {\bibfnamefont {E.}~\bibnamefont {Khatami}}, \bibinfo {author} {\bibfnamefont {E.}~\bibnamefont {Demler}},\ and\ \bibinfo {author} {\bibfnamefont {M.}~\bibnamefont {Greiner}},\ }\bibfield  {title} {\bibinfo {title} {Observation of nagaoka polarons in a fermi--hubbard quantum simulator},\ }\href {https://doi.org/10.1038/s41586-024-07272-9} {\bibfield  {journal} {\bibinfo  {journal} {Nature}\ }\textbf {\bibinfo {volume} {629}},\ \bibinfo {pages} {317} (\bibinfo {year} {2024})}\BibitemShut {NoStop}%
\bibitem [{\citenamefont {Prichard}\ \emph {et~al.}(2024)\citenamefont {Prichard}, \citenamefont {Spar}, \citenamefont {Morera}, \citenamefont {Demler}, \citenamefont {Yan},\ and\ \citenamefont {Bakr}}]{prichard2023directly}%
  \BibitemOpen
  \bibfield  {author} {\bibinfo {author} {\bibfnamefont {M.~L.}\ \bibnamefont {Prichard}}, \bibinfo {author} {\bibfnamefont {B.~M.}\ \bibnamefont {Spar}}, \bibinfo {author} {\bibfnamefont {I.}~\bibnamefont {Morera}}, \bibinfo {author} {\bibfnamefont {E.}~\bibnamefont {Demler}}, \bibinfo {author} {\bibfnamefont {Z.~Z.}\ \bibnamefont {Yan}},\ and\ \bibinfo {author} {\bibfnamefont {W.~S.}\ \bibnamefont {Bakr}},\ }\bibfield  {title} {\bibinfo {title} {{Directly imaging spin polarons in a kinetically frustrated Hubbard system}},\ }\href {https://doi.org/10.1038/s41586-024-07356-6} {\bibfield  {journal} {\bibinfo  {journal} {Nature}\ }\textbf {\bibinfo {volume} {629}},\ \bibinfo {pages} {323} (\bibinfo {year} {2024})}\BibitemShut {NoStop}%
\bibitem [{\citenamefont {Singh}\ and\ \citenamefont {Huse}(1992)}]{SinghHuse}%
  \BibitemOpen
  \bibfield  {author} {\bibinfo {author} {\bibfnamefont {R.~R.~P.}\ \bibnamefont {Singh}}\ and\ \bibinfo {author} {\bibfnamefont {D.~A.}\ \bibnamefont {Huse}},\ }\bibfield  {title} {\bibinfo {title} {{Three-sublattice order in triangular- and Kagom\'e-lattice spin-half antiferromagnets}},\ }\href {https://doi.org/10.1103/PhysRevLett.68.1766} {\bibfield  {journal} {\bibinfo  {journal} {Phys. Rev. Lett.}\ }\textbf {\bibinfo {volume} {68}},\ \bibinfo {pages} {1766} (\bibinfo {year} {1992})}\BibitemShut {NoStop}%
\bibitem [{\citenamefont {Hanisch}\ \emph {et~al.}(1995)\citenamefont {Hanisch}, \citenamefont {Kleine}, \citenamefont {Ritzl},\ and\ \citenamefont {Müller-Hartmann}}]{ANDP}%
  \BibitemOpen
  \bibfield  {author} {\bibinfo {author} {\bibfnamefont {T.}~\bibnamefont {Hanisch}}, \bibinfo {author} {\bibfnamefont {B.}~\bibnamefont {Kleine}}, \bibinfo {author} {\bibfnamefont {A.}~\bibnamefont {Ritzl}},\ and\ \bibinfo {author} {\bibfnamefont {E.}~\bibnamefont {Müller-Hartmann}},\ }\bibfield  {title} {\bibinfo {title} {{Ferromagnetism in the Hubbard model: instability of the Nagaoka state on the triangular, honeycomb and kagome lattices}},\ }\href {https://doi.org/https://doi.org/10.1002/andp.19955070405} {\bibfield  {journal} {\bibinfo  {journal} {Annalen der Physik}\ }\textbf {\bibinfo {volume} {507}},\ \bibinfo {pages} {303} (\bibinfo {year} {1995})}\BibitemShut {NoStop}%
\bibitem [{\citenamefont {Haerter}\ and\ \citenamefont {Shastry}(2005)}]{KineticAFM}%
  \BibitemOpen
  \bibfield  {author} {\bibinfo {author} {\bibfnamefont {J.~O.}\ \bibnamefont {Haerter}}\ and\ \bibinfo {author} {\bibfnamefont {B.~S.}\ \bibnamefont {Shastry}},\ }\bibfield  {title} {\bibinfo {title} {{Kinetic Antiferromagnetism in the Triangular Lattice}},\ }\href {https://doi.org/10.1103/PhysRevLett.95.087202} {\bibfield  {journal} {\bibinfo  {journal} {Phys. Rev. Lett.}\ }\textbf {\bibinfo {volume} {95}},\ \bibinfo {pages} {087202} (\bibinfo {year} {2005})}\BibitemShut {NoStop}%
\bibitem [{\citenamefont {Li}\ \emph {et~al.}(2022)\citenamefont {Li}, \citenamefont {Li}, \citenamefont {Zhao}, \citenamefont {Luo},\ and\ \citenamefont {Xie}}]{Li2dtensor}%
  \BibitemOpen
  \bibfield  {author} {\bibinfo {author} {\bibfnamefont {Q.}~\bibnamefont {Li}}, \bibinfo {author} {\bibfnamefont {H.}~\bibnamefont {Li}}, \bibinfo {author} {\bibfnamefont {J.}~\bibnamefont {Zhao}}, \bibinfo {author} {\bibfnamefont {H.-G.}\ \bibnamefont {Luo}},\ and\ \bibinfo {author} {\bibfnamefont {Z.~Y.}\ \bibnamefont {Xie}},\ }\bibfield  {title} {\bibinfo {title} {{Magnetization of the spin-$\frac{1}{2}$ Heisenberg antiferromagnet on the triangular lattice}},\ }\href {https://doi.org/10.1103/PhysRevB.105.184418} {\bibfield  {journal} {\bibinfo  {journal} {Phys. Rev. B}\ }\textbf {\bibinfo {volume} {105}},\ \bibinfo {pages} {184418} (\bibinfo {year} {2022})}\BibitemShut {NoStop}%
\bibitem [{\citenamefont {Zhu}\ and\ \citenamefont {White}(2015)}]{ZhuTriangular}%
  \BibitemOpen
  \bibfield  {author} {\bibinfo {author} {\bibfnamefont {Z.}~\bibnamefont {Zhu}}\ and\ \bibinfo {author} {\bibfnamefont {S.~R.}\ \bibnamefont {White}},\ }\bibfield  {title} {\bibinfo {title} {{Spin liquid phase of the $S = \frac{1}{2}$ ${J}_{1} \ensuremath{-} {J}_{2}$ Heisenberg model on the triangular lattice}},\ }\href {https://doi.org/10.1103/PhysRevB.92.041105} {\bibfield  {journal} {\bibinfo  {journal} {Physical Review B}\ }\textbf {\bibinfo {volume} {92}},\ \bibinfo {pages} {041105} (\bibinfo {year} {2015})}\BibitemShut {NoStop}%
\bibitem [{\citenamefont {Jiang}\ and\ \citenamefont {Jiang}(2023)}]{JiangTriangular}%
  \BibitemOpen
  \bibfield  {author} {\bibinfo {author} {\bibfnamefont {Y.-F.}\ \bibnamefont {Jiang}}\ and\ \bibinfo {author} {\bibfnamefont {H.-C.}\ \bibnamefont {Jiang}},\ }\bibfield  {title} {\bibinfo {title} {{Nature of quantum spin liquids of the $S=\frac{1}{2}$ Heisenberg antiferromagnet on the triangular lattice: A parallel DMRG study}},\ }\href {https://doi.org/10.1103/PhysRevB.107.L140411} {\bibfield  {journal} {\bibinfo  {journal} {Phys. Rev. B}\ }\textbf {\bibinfo {volume} {107}},\ \bibinfo {pages} {L140411} (\bibinfo {year} {2023})}\BibitemShut {NoStop}%
\bibitem [{\citenamefont {Pan}\ \emph {et~al.}(2020{\natexlab{a}})\citenamefont {Pan}, \citenamefont {Wu},\ and\ \citenamefont {Das~Sarma}}]{PanWu1}%
  \BibitemOpen
  \bibfield  {author} {\bibinfo {author} {\bibfnamefont {H.}~\bibnamefont {Pan}}, \bibinfo {author} {\bibfnamefont {F.}~\bibnamefont {Wu}},\ and\ \bibinfo {author} {\bibfnamefont {S.}~\bibnamefont {Das~Sarma}},\ }\bibfield  {title} {\bibinfo {title} {{Quantum phase diagram of a Moir\'e-Hubbard model}},\ }\href {https://doi.org/10.1103/PhysRevB.102.201104} {\bibfield  {journal} {\bibinfo  {journal} {Phys. Rev. B}\ }\textbf {\bibinfo {volume} {102}},\ \bibinfo {pages} {201104} (\bibinfo {year} {2020}{\natexlab{a}})}\BibitemShut {NoStop}%
\bibitem [{\citenamefont {Kiese}\ \emph {et~al.}(2022)\citenamefont {Kiese}, \citenamefont {He}, \citenamefont {Hickey}, \citenamefont {Rubio},\ and\ \citenamefont {Kennes}}]{APLM}%
  \BibitemOpen
  \bibfield  {author} {\bibinfo {author} {\bibfnamefont {D.}~\bibnamefont {Kiese}}, \bibinfo {author} {\bibfnamefont {Y.}~\bibnamefont {He}}, \bibinfo {author} {\bibfnamefont {C.}~\bibnamefont {Hickey}}, \bibinfo {author} {\bibfnamefont {A.}~\bibnamefont {Rubio}},\ and\ \bibinfo {author} {\bibfnamefont {D.~M.}\ \bibnamefont {Kennes}},\ }\bibfield  {title} {\bibinfo {title} {{TMDs as a platform for spin liquid physics: A strong coupling study of twisted bilayer $\mathrm{WSe}_{2}$}},\ }\href {https://doi.org/10.1063/5.0077901} {\bibfield  {journal} {\bibinfo  {journal} {APL Materials}\ }\textbf {\bibinfo {volume} {10}},\ \bibinfo {pages} {031113} (\bibinfo {year} {2022})}\BibitemShut {NoStop}%
\bibitem [{\citenamefont {Peng}\ \emph {et~al.}(2021)\citenamefont {Peng}, \citenamefont {Jiang}, \citenamefont {Wang},\ and\ \citenamefont {Jiang}}]{PengPDW}%
  \BibitemOpen
  \bibfield  {author} {\bibinfo {author} {\bibfnamefont {C.}~\bibnamefont {Peng}}, \bibinfo {author} {\bibfnamefont {Y.-F.}\ \bibnamefont {Jiang}}, \bibinfo {author} {\bibfnamefont {Y.}~\bibnamefont {Wang}},\ and\ \bibinfo {author} {\bibfnamefont {H.-C.}\ \bibnamefont {Jiang}},\ }\bibfield  {title} {\bibinfo {title} {{Gapless spin liquid and pair density wave of the Hubbard model on three-leg triangular cylinders}},\ }\href {https://doi.org/10.1088/1367-2630/ac3a83} {\bibfield  {journal} {\bibinfo  {journal} {New Journal of Physics}\ }\textbf {\bibinfo {volume} {23}},\ \bibinfo {pages} {123004} (\bibinfo {year} {2021})}\BibitemShut {NoStop}%
\bibitem [{\citenamefont {Wietek}\ \emph {et~al.}(2022)\citenamefont {Wietek}, \citenamefont {Wang}, \citenamefont {Zang}, \citenamefont {Cano}, \citenamefont {Georges},\ and\ \citenamefont {Millis}}]{WietekWang}%
  \BibitemOpen
  \bibfield  {author} {\bibinfo {author} {\bibfnamefont {A.}~\bibnamefont {Wietek}}, \bibinfo {author} {\bibfnamefont {J.}~\bibnamefont {Wang}}, \bibinfo {author} {\bibfnamefont {J.}~\bibnamefont {Zang}}, \bibinfo {author} {\bibfnamefont {J.}~\bibnamefont {Cano}}, \bibinfo {author} {\bibfnamefont {A.}~\bibnamefont {Georges}},\ and\ \bibinfo {author} {\bibfnamefont {A.}~\bibnamefont {Millis}},\ }\bibfield  {title} {\bibinfo {title} {{Tunable stripe order and weak superconductivity in the Moir{\'{e} } Hubbard model}},\ }\bibfield  {journal} {\bibinfo  {journal} {Physical Review Research}\ }\textbf {\bibinfo {volume} {4}},\ \href {https://doi.org/10.1103/physrevresearch.4.043048} {10.1103/physrevresearch.4.043048} (\bibinfo {year} {2022})\BibitemShut {NoStop}%
\bibitem [{\citenamefont {Zhu}\ \emph {et~al.}(2022)\citenamefont {Zhu}, \citenamefont {Sheng},\ and\ \citenamefont {Vishwanath}}]{Zhu}%
  \BibitemOpen
  \bibfield  {author} {\bibinfo {author} {\bibfnamefont {Z.}~\bibnamefont {Zhu}}, \bibinfo {author} {\bibfnamefont {D.~N.}\ \bibnamefont {Sheng}},\ and\ \bibinfo {author} {\bibfnamefont {A.}~\bibnamefont {Vishwanath}},\ }\bibfield  {title} {\bibinfo {title} {{Doped Mott insulators in the triangular-lattice Hubbard model}},\ }\href {https://doi.org/10.1103/PhysRevB.105.205110} {\bibfield  {journal} {\bibinfo  {journal} {Phys. Rev. B}\ }\textbf {\bibinfo {volume} {105}},\ \bibinfo {pages} {205110} (\bibinfo {year} {2022})}\BibitemShut {NoStop}%
\bibitem [{\citenamefont {B\'elanger}\ \emph {et~al.}(2022)\citenamefont {B\'elanger}, \citenamefont {Fournier},\ and\ \citenamefont {S\'en\'echal}}]{PhysRevB.106.235135}%
  \BibitemOpen
  \bibfield  {author} {\bibinfo {author} {\bibfnamefont {M.}~\bibnamefont {B\'elanger}}, \bibinfo {author} {\bibfnamefont {J.}~\bibnamefont {Fournier}},\ and\ \bibinfo {author} {\bibfnamefont {D.}~\bibnamefont {S\'en\'echal}},\ }\bibfield  {title} {\bibinfo {title} {{Superconductivity in the twisted bilayer transition metal dichalcogenide ${\mathrm{WSe}}_{2}$: A quantum cluster study}},\ }\href {https://doi.org/10.1103/PhysRevB.106.235135} {\bibfield  {journal} {\bibinfo  {journal} {Phys. Rev. B}\ }\textbf {\bibinfo {volume} {106}},\ \bibinfo {pages} {235135} (\bibinfo {year} {2022})}\BibitemShut {NoStop}%
\bibitem [{\citenamefont {Chen}\ and\ \citenamefont {Sheng}(2023)}]{chen2023singlet}%
  \BibitemOpen
  \bibfield  {author} {\bibinfo {author} {\bibfnamefont {F.}~\bibnamefont {Chen}}\ and\ \bibinfo {author} {\bibfnamefont {D.~N.}\ \bibnamefont {Sheng}},\ }\bibfield  {title} {\bibinfo {title} {Singlet, triplet, and pair density wave superconductivity in the doped triangular-lattice moir\'e system},\ }\href {https://doi.org/10.1103/PhysRevB.108.L201110} {\bibfield  {journal} {\bibinfo  {journal} {Phys. Rev. B}\ }\textbf {\bibinfo {volume} {108}},\ \bibinfo {pages} {L201110} (\bibinfo {year} {2023})}\BibitemShut {NoStop}%
\bibitem [{\citenamefont {Zegrodnik}\ and\ \citenamefont {Biborski}(2023)}]{zegrodnik2023mixed}%
  \BibitemOpen
  \bibfield  {author} {\bibinfo {author} {\bibfnamefont {M.}~\bibnamefont {Zegrodnik}}\ and\ \bibinfo {author} {\bibfnamefont {A.}~\bibnamefont {Biborski}},\ }\bibfield  {title} {\bibinfo {title} {Mixed singlet-triplet superconducting state within the moir\'e $t\text{\ensuremath{-}}j\text{\ensuremath{-}}u$ model applied to twisted bilayer ${\mathrm{wse}}_{2}$},\ }\href {https://doi.org/10.1103/PhysRevB.108.064506} {\bibfield  {journal} {\bibinfo  {journal} {Phys. Rev. B}\ }\textbf {\bibinfo {volume} {108}},\ \bibinfo {pages} {064506} (\bibinfo {year} {2023})}\BibitemShut {NoStop}%
\bibitem [{\citenamefont {Honerkamp}(2003)}]{Honerkamp}%
  \BibitemOpen
  \bibfield  {author} {\bibinfo {author} {\bibfnamefont {C.}~\bibnamefont {Honerkamp}},\ }\bibfield  {title} {\bibinfo {title} {{Instabilities of interacting electrons on the triangular lattice}},\ }\href {https://doi.org/10.1103/PhysRevB.68.104510} {\bibfield  {journal} {\bibinfo  {journal} {Phys. Rev. B}\ }\textbf {\bibinfo {volume} {68}},\ \bibinfo {pages} {104510} (\bibinfo {year} {2003})}\BibitemShut {NoStop}%
\bibitem [{\citenamefont {Martin}\ and\ \citenamefont {Batista}(2008)}]{MartinBatista}%
  \BibitemOpen
  \bibfield  {author} {\bibinfo {author} {\bibfnamefont {I.}~\bibnamefont {Martin}}\ and\ \bibinfo {author} {\bibfnamefont {C.~D.}\ \bibnamefont {Batista}},\ }\bibfield  {title} {\bibinfo {title} {{Itinerant Electron-Driven Chiral Magnetic Ordering and Spontaneous Quantum Hall Effect in Triangular Lattice Models}},\ }\href {https://doi.org/10.1103/PhysRevLett.101.156402} {\bibfield  {journal} {\bibinfo  {journal} {Phys. Rev. Lett.}\ }\textbf {\bibinfo {volume} {101}},\ \bibinfo {pages} {156402} (\bibinfo {year} {2008})}\BibitemShut {NoStop}%
\bibitem [{\citenamefont {Li}(2010)}]{li2010metalinsulator}%
  \BibitemOpen
  \bibfield  {author} {\bibinfo {author} {\bibfnamefont {T.}~\bibnamefont {Li}},\ }\href@noop {} {\bibinfo {title} {{Metal-insulator transition in quarter-filled Hubbard model on triangular lattice and its implication for the physics of $\mathrm{Na}_{0.5}\mathrm{CoO}_{2}$}}} (\bibinfo {year} {2010}),\ \Eprint {https://arxiv.org/abs/1001.0620} {arXiv:1001.0620 [cond-mat.str-el]} \BibitemShut {NoStop}%
\bibitem [{\citenamefont {Ye}\ \emph {et~al.}(2016)\citenamefont {Ye}, \citenamefont {Mesaros},\ and\ \citenamefont {Ran}}]{YeRan}%
  \BibitemOpen
  \bibfield  {author} {\bibinfo {author} {\bibfnamefont {B.}~\bibnamefont {Ye}}, \bibinfo {author} {\bibfnamefont {A.}~\bibnamefont {Mesaros}},\ and\ \bibinfo {author} {\bibfnamefont {Y.}~\bibnamefont {Ran}},\ }\href@noop {} {\bibinfo {title} {{Ferromagnetism and d+id superconductivity in 1/2 doped correlated systems on triangular lattice}}} (\bibinfo {year} {2016}),\ \Eprint {https://arxiv.org/abs/1604.08615} {arXiv:1604.08615 [cond-mat.str-el]} \BibitemShut {NoStop}%
\bibitem [{\citenamefont {Pasrija}\ and\ \citenamefont {Kumar}(2016)}]{PasrijaKumar}%
  \BibitemOpen
  \bibfield  {author} {\bibinfo {author} {\bibfnamefont {K.}~\bibnamefont {Pasrija}}\ and\ \bibinfo {author} {\bibfnamefont {S.}~\bibnamefont {Kumar}},\ }\bibfield  {title} {\bibinfo {title} {{Noncollinear and noncoplanar magnetic order in the extended Hubbard model on anisotropic triangular lattice}},\ }\href {https://doi.org/10.1103/PhysRevB.93.195110} {\bibfield  {journal} {\bibinfo  {journal} {Phys. Rev. B}\ }\textbf {\bibinfo {volume} {93}},\ \bibinfo {pages} {195110} (\bibinfo {year} {2016})}\BibitemShut {NoStop}%
\bibitem [{\citenamefont {Pan}\ \emph {et~al.}(2020{\natexlab{b}})\citenamefont {Pan}, \citenamefont {Wu},\ and\ \citenamefont {Das~Sarma}}]{PanWu2}%
  \BibitemOpen
  \bibfield  {author} {\bibinfo {author} {\bibfnamefont {H.}~\bibnamefont {Pan}}, \bibinfo {author} {\bibfnamefont {F.}~\bibnamefont {Wu}},\ and\ \bibinfo {author} {\bibfnamefont {S.}~\bibnamefont {Das~Sarma}},\ }\bibfield  {title} {\bibinfo {title} {{Band topology, Hubbard model, Heisenberg model, and Dzyaloshinskii-Moriya interaction in twisted bilayer ${\mathrm{WSe}}_{2}$}},\ }\href {https://doi.org/10.1103/PhysRevResearch.2.033087} {\bibfield  {journal} {\bibinfo  {journal} {Phys. Rev. Res.}\ }\textbf {\bibinfo {volume} {2}},\ \bibinfo {pages} {033087} (\bibinfo {year} {2020}{\natexlab{b}})}\BibitemShut {NoStop}%
\bibitem [{\citenamefont {Zang}\ \emph {et~al.}(2021)\citenamefont {Zang}, \citenamefont {Wang}, \citenamefont {Cano},\ and\ \citenamefont {Millis}}]{ZangWang}%
  \BibitemOpen
  \bibfield  {author} {\bibinfo {author} {\bibfnamefont {J.}~\bibnamefont {Zang}}, \bibinfo {author} {\bibfnamefont {J.}~\bibnamefont {Wang}}, \bibinfo {author} {\bibfnamefont {J.}~\bibnamefont {Cano}},\ and\ \bibinfo {author} {\bibfnamefont {A.~J.}\ \bibnamefont {Millis}},\ }\bibfield  {title} {\bibinfo {title} {{Hartree-Fock study of the moir\'e Hubbard model for twisted bilayer transition metal dichalcogenides}},\ }\href {https://doi.org/10.1103/PhysRevB.104.075150} {\bibfield  {journal} {\bibinfo  {journal} {Phys. Rev. B}\ }\textbf {\bibinfo {volume} {104}},\ \bibinfo {pages} {075150} (\bibinfo {year} {2021})}\BibitemShut {NoStop}%
\bibitem [{\citenamefont {Classen}\ \emph {et~al.}(2019)\citenamefont {Classen}, \citenamefont {Honerkamp},\ and\ \citenamefont {Scherer}}]{PhysRevB.99.195120}%
  \BibitemOpen
  \bibfield  {author} {\bibinfo {author} {\bibfnamefont {L.}~\bibnamefont {Classen}}, \bibinfo {author} {\bibfnamefont {C.}~\bibnamefont {Honerkamp}},\ and\ \bibinfo {author} {\bibfnamefont {M.~M.}\ \bibnamefont {Scherer}},\ }\bibfield  {title} {\bibinfo {title} {{Competing phases of interacting electrons on triangular lattices in moir\'e heterostructures}},\ }\href {https://doi.org/10.1103/PhysRevB.99.195120} {\bibfield  {journal} {\bibinfo  {journal} {Phys. Rev. B}\ }\textbf {\bibinfo {volume} {99}},\ \bibinfo {pages} {195120} (\bibinfo {year} {2019})}\BibitemShut {NoStop}%
\bibitem [{\citenamefont {Scherer}\ \emph {et~al.}(2022)\citenamefont {Scherer}, \citenamefont {Kennes},\ and\ \citenamefont {Classen}}]{SchererKennes}%
  \BibitemOpen
  \bibfield  {author} {\bibinfo {author} {\bibfnamefont {M.~M.}\ \bibnamefont {Scherer}}, \bibinfo {author} {\bibfnamefont {D.~M.}\ \bibnamefont {Kennes}},\ and\ \bibinfo {author} {\bibfnamefont {L.}~\bibnamefont {Classen}},\ }\bibfield  {title} {\bibinfo {title} {{Chiral superconductivity with enhanced quantized Hall responses in moir{\'e} transition metal dichalcogenides}},\ }\href {https://doi.org/10.1038/s41535-022-00504-z} {\bibfield  {journal} {\bibinfo  {journal} {npj Quantum Materials}\ }\textbf {\bibinfo {volume} {7}},\ \bibinfo {pages} {100} (\bibinfo {year} {2022})}\BibitemShut {NoStop}%
\bibitem [{\citenamefont {Wu}\ \emph {et~al.}(2023)\citenamefont {Wu}, \citenamefont {Wu},\ and\ \citenamefont {Yao}}]{Yao}%
  \BibitemOpen
  \bibfield  {author} {\bibinfo {author} {\bibfnamefont {Y.-M.}\ \bibnamefont {Wu}}, \bibinfo {author} {\bibfnamefont {Z.}~\bibnamefont {Wu}},\ and\ \bibinfo {author} {\bibfnamefont {H.}~\bibnamefont {Yao}},\ }\bibfield  {title} {\bibinfo {title} {{Pair-Density-Wave and Chiral Superconductivity in Twisted Bilayer Transition Metal Dichalcogenides}},\ }\href {https://doi.org/10.1103/PhysRevLett.130.126001} {\bibfield  {journal} {\bibinfo  {journal} {Phys. Rev. Lett.}\ }\textbf {\bibinfo {volume} {130}},\ \bibinfo {pages} {126001} (\bibinfo {year} {2023})}\BibitemShut {NoStop}%
\bibitem [{\citenamefont {Li}\ \emph {et~al.}(2014)\citenamefont {Li}, \citenamefont {Antipov}, \citenamefont {Rubtsov}, \citenamefont {Kirchner},\ and\ \citenamefont {Hanke}}]{DMFTTLHM}%
  \BibitemOpen
  \bibfield  {author} {\bibinfo {author} {\bibfnamefont {G.}~\bibnamefont {Li}}, \bibinfo {author} {\bibfnamefont {A.~E.}\ \bibnamefont {Antipov}}, \bibinfo {author} {\bibfnamefont {A.~N.}\ \bibnamefont {Rubtsov}}, \bibinfo {author} {\bibfnamefont {S.}~\bibnamefont {Kirchner}},\ and\ \bibinfo {author} {\bibfnamefont {W.}~\bibnamefont {Hanke}},\ }\bibfield  {title} {\bibinfo {title} {{Competing phases of the Hubbard model on a triangular lattice: Insights from the entropy}},\ }\href {https://doi.org/10.1103/PhysRevB.89.161118} {\bibfield  {journal} {\bibinfo  {journal} {Phys. Rev. B}\ }\textbf {\bibinfo {volume} {89}},\ \bibinfo {pages} {161118} (\bibinfo {year} {2014})}\BibitemShut {NoStop}%
\bibitem [{\citenamefont {Potasz}\ \emph {et~al.}(2024)\citenamefont {Potasz}, \citenamefont {Morales-Dur\'an}, \citenamefont {Hu},\ and\ \citenamefont {MacDonald}}]{potasz2023itinerant}%
  \BibitemOpen
  \bibfield  {author} {\bibinfo {author} {\bibfnamefont {P.}~\bibnamefont {Potasz}}, \bibinfo {author} {\bibfnamefont {N.}~\bibnamefont {Morales-Dur\'an}}, \bibinfo {author} {\bibfnamefont {N.~C.}\ \bibnamefont {Hu}},\ and\ \bibinfo {author} {\bibfnamefont {A.~H.}\ \bibnamefont {MacDonald}},\ }\bibfield  {title} {\bibinfo {title} {Itinerant ferromagnetism in transition metal dichalcogenide moir\'e superlattices},\ }\href {https://doi.org/10.1103/PhysRevB.109.045144} {\bibfield  {journal} {\bibinfo  {journal} {Phys. Rev. B}\ }\textbf {\bibinfo {volume} {109}},\ \bibinfo {pages} {045144} (\bibinfo {year} {2024})}\BibitemShut {NoStop}%
\bibitem [{\citenamefont {Li}(2012)}]{Li_2012}%
  \BibitemOpen
  \bibfield  {author} {\bibinfo {author} {\bibfnamefont {T.}~\bibnamefont {Li}},\ }\bibfield  {title} {\bibinfo {title} {{Spontaneous quantum Hall effect in quarter-doped Hubbard model on honeycomb lattice and its possible realization in doped graphene system}},\ }\href {https://doi.org/10.1209/0295-5075/97/37001} {\bibfield  {journal} {\bibinfo  {journal} {Europhysics Letters}\ }\textbf {\bibinfo {volume} {97}},\ \bibinfo {pages} {37001} (\bibinfo {year} {2012})}\BibitemShut {NoStop}%
\bibitem [{\citenamefont {Wang}\ \emph {et~al.}(2012)\citenamefont {Wang}, \citenamefont {Xiang}, \citenamefont {Wang}, \citenamefont {Wang}, \citenamefont {Yang},\ and\ \citenamefont {Lee}}]{WangXiang}%
  \BibitemOpen
  \bibfield  {author} {\bibinfo {author} {\bibfnamefont {W.-S.}\ \bibnamefont {Wang}}, \bibinfo {author} {\bibfnamefont {Y.-Y.}\ \bibnamefont {Xiang}}, \bibinfo {author} {\bibfnamefont {Q.-H.}\ \bibnamefont {Wang}}, \bibinfo {author} {\bibfnamefont {F.}~\bibnamefont {Wang}}, \bibinfo {author} {\bibfnamefont {F.}~\bibnamefont {Yang}},\ and\ \bibinfo {author} {\bibfnamefont {D.-H.}\ \bibnamefont {Lee}},\ }\bibfield  {title} {\bibinfo {title} {{Functional renormalization group and variational Monte Carlo studies of the electronic instabilities in graphene near $\frac{1}{4}$ doping}},\ }\href {https://doi.org/10.1103/PhysRevB.85.035414} {\bibfield  {journal} {\bibinfo  {journal} {Phys. Rev. B}\ }\textbf {\bibinfo {volume} {85}},\ \bibinfo {pages} {035414} (\bibinfo {year} {2012})}\BibitemShut {NoStop}%
\bibitem [{\citenamefont {Nandkishore}\ \emph {et~al.}(2012{\natexlab{a}})\citenamefont {Nandkishore}, \citenamefont {Levitov},\ and\ \citenamefont {Chubukov}}]{NandkishoreLevitov}%
  \BibitemOpen
  \bibfield  {author} {\bibinfo {author} {\bibfnamefont {R.}~\bibnamefont {Nandkishore}}, \bibinfo {author} {\bibfnamefont {L.~S.}\ \bibnamefont {Levitov}},\ and\ \bibinfo {author} {\bibfnamefont {A.~V.}\ \bibnamefont {Chubukov}},\ }\bibfield  {title} {\bibinfo {title} {{Chiral superconductivity from repulsive interactions in doped graphene}},\ }\href {https://doi.org/10.1038/nphys2208} {\bibfield  {journal} {\bibinfo  {journal} {Nature Physics}\ }\textbf {\bibinfo {volume} {8}},\ \bibinfo {pages} {158} (\bibinfo {year} {2012}{\natexlab{a}})}\BibitemShut {NoStop}%
\bibitem [{\citenamefont {Nandkishore}\ \emph {et~al.}(2012{\natexlab{b}})\citenamefont {Nandkishore}, \citenamefont {Chern},\ and\ \citenamefont {Chubukov}}]{NandkishorChern}%
  \BibitemOpen
  \bibfield  {author} {\bibinfo {author} {\bibfnamefont {R.}~\bibnamefont {Nandkishore}}, \bibinfo {author} {\bibfnamefont {G.-W.}\ \bibnamefont {Chern}},\ and\ \bibinfo {author} {\bibfnamefont {A.~V.}\ \bibnamefont {Chubukov}},\ }\bibfield  {title} {\bibinfo {title} {{Itinerant Half-Metal Spin-Density-Wave State on the Hexagonal Lattice}},\ }\href {https://doi.org/10.1103/PhysRevLett.108.227204} {\bibfield  {journal} {\bibinfo  {journal} {Phys. Rev. Lett.}\ }\textbf {\bibinfo {volume} {108}},\ \bibinfo {pages} {227204} (\bibinfo {year} {2012}{\natexlab{b}})}\BibitemShut {NoStop}%
\bibitem [{\citenamefont {Jiang}\ \emph {et~al.}(2014)\citenamefont {Jiang}, \citenamefont {Mesaros},\ and\ \citenamefont {Ran}}]{JiangMesaros}%
  \BibitemOpen
  \bibfield  {author} {\bibinfo {author} {\bibfnamefont {S.}~\bibnamefont {Jiang}}, \bibinfo {author} {\bibfnamefont {A.}~\bibnamefont {Mesaros}},\ and\ \bibinfo {author} {\bibfnamefont {Y.}~\bibnamefont {Ran}},\ }\bibfield  {title} {\bibinfo {title} {{Chiral Spin-Density Wave, Spin-Charge-Chern Liquid, and $d+id$ Superconductivity in $1/4$-Doped Correlated Electronic Systems on the Honeycomb Lattice}},\ }\href {https://doi.org/10.1103/PhysRevX.4.031040} {\bibfield  {journal} {\bibinfo  {journal} {Phys. Rev. X}\ }\textbf {\bibinfo {volume} {4}},\ \bibinfo {pages} {031040} (\bibinfo {year} {2014})}\BibitemShut {NoStop}%
\bibitem [{\citenamefont {Akagi}\ and\ \citenamefont {Motome}(2010)}]{AkagiMotome}%
  \BibitemOpen
  \bibfield  {author} {\bibinfo {author} {\bibfnamefont {Y.}~\bibnamefont {Akagi}}\ and\ \bibinfo {author} {\bibfnamefont {Y.}~\bibnamefont {Motome}},\ }\bibfield  {title} {\bibinfo {title} {{Spin Chirality Ordering and Anomalous Hall Effect in the Ferromagnetic Kondo Lattice Model on a Triangular Lattice}},\ }\href {https://doi.org/10.1143/JPSJ.79.083711} {\bibfield  {journal} {\bibinfo  {journal} {Journal of the Physical Society of Japan}\ }\textbf {\bibinfo {volume} {79}},\ \bibinfo {pages} {083711} (\bibinfo {year} {2010})}\BibitemShut {NoStop}%
\bibitem [{\citenamefont {Cancès}\ and\ \citenamefont {Le~Bris}(2000)}]{ODA}%
  \BibitemOpen
  \bibfield  {author} {\bibinfo {author} {\bibfnamefont {E.}~\bibnamefont {Cancès}}\ and\ \bibinfo {author} {\bibfnamefont {C.}~\bibnamefont {Le~Bris}},\ }\bibfield  {title} {\bibinfo {title} {{Can we outperform the DIIS approach for electronic structure calculations?}},\ }\href {https://doi.org/https://doi.org/10.1002/1097-461X(2000)79:2<82::AID-QUA3>3.0.CO;2-I} {\bibfield  {journal} {\bibinfo  {journal} {International Journal of Quantum Chemistry}\ }\textbf {\bibinfo {volume} {79}},\ \bibinfo {pages} {82} (\bibinfo {year} {2000})}\BibitemShut {NoStop}%
\bibitem [{\citenamefont {White}(1992)}]{White}%
  \BibitemOpen
  \bibfield  {author} {\bibinfo {author} {\bibfnamefont {S.~R.}\ \bibnamefont {White}},\ }\bibfield  {title} {\bibinfo {title} {{Density matrix formulation for quantum renormalization groups}},\ }\href {https://doi.org/10.1103/PhysRevLett.69.2863} {\bibfield  {journal} {\bibinfo  {journal} {Phys. Rev. Lett.}\ }\textbf {\bibinfo {volume} {69}},\ \bibinfo {pages} {2863} (\bibinfo {year} {1992})}\BibitemShut {NoStop}%
\bibitem [{\citenamefont {Hubig}\ \emph {et~al.}(2015)\citenamefont {Hubig}, \citenamefont {McCulloch}, \citenamefont {Schollw\"ock},\ and\ \citenamefont {Wolf}}]{SingleSite}%
  \BibitemOpen
  \bibfield  {author} {\bibinfo {author} {\bibfnamefont {C.}~\bibnamefont {Hubig}}, \bibinfo {author} {\bibfnamefont {I.~P.}\ \bibnamefont {McCulloch}}, \bibinfo {author} {\bibfnamefont {U.}~\bibnamefont {Schollw\"ock}},\ and\ \bibinfo {author} {\bibfnamefont {F.~A.}\ \bibnamefont {Wolf}},\ }\bibfield  {title} {\bibinfo {title} {{Strictly single-site DMRG algorithm with subspace expansion}},\ }\href {https://doi.org/10.1103/PhysRevB.91.155115} {\bibfield  {journal} {\bibinfo  {journal} {Phys. Rev. B}\ }\textbf {\bibinfo {volume} {91}},\ \bibinfo {pages} {155115} (\bibinfo {year} {2015})}\BibitemShut {NoStop}%
\bibitem [{\citenamefont {Verstraete}\ and\ \citenamefont {Cirac}(2004)}]{verstraete2004}%
  \BibitemOpen
  \bibfield  {author} {\bibinfo {author} {\bibfnamefont {F.}~\bibnamefont {Verstraete}}\ and\ \bibinfo {author} {\bibfnamefont {J.~I.}\ \bibnamefont {Cirac}},\ }\bibfield  {title} {\bibinfo {title} {{Renormalization algorithms for quantum-many body systems in two and higher dimensions}},\ }\href@noop {} {\bibfield  {journal} {\bibinfo  {journal} {arXiv preprint cond-mat/0407066}\ } (\bibinfo {year} {2004})}\BibitemShut {NoStop}%
\bibitem [{\citenamefont {Nishio}\ \emph {et~al.}(2004)\citenamefont {Nishio}, \citenamefont {Maeshima}, \citenamefont {Gendiar},\ and\ \citenamefont {Nishino}}]{nishio2004tensor}%
  \BibitemOpen
  \bibfield  {author} {\bibinfo {author} {\bibfnamefont {Y.}~\bibnamefont {Nishio}}, \bibinfo {author} {\bibfnamefont {N.}~\bibnamefont {Maeshima}}, \bibinfo {author} {\bibfnamefont {A.}~\bibnamefont {Gendiar}},\ and\ \bibinfo {author} {\bibfnamefont {T.}~\bibnamefont {Nishino}},\ }\bibfield  {title} {\bibinfo {title} {{Tensor product variational formulation for quantum systems}},\ }\href@noop {} {\bibfield  {journal} {\bibinfo  {journal} {arXiv preprint cond-mat/0401115}\ } (\bibinfo {year} {2004})}\BibitemShut {NoStop}%
\bibitem [{\citenamefont {Nishino}\ and\ \citenamefont {Okunishi}(1996)}]{CTM}%
  \BibitemOpen
  \bibfield  {author} {\bibinfo {author} {\bibfnamefont {T.}~\bibnamefont {Nishino}}\ and\ \bibinfo {author} {\bibfnamefont {K.}~\bibnamefont {Okunishi}},\ }\bibfield  {title} {\bibinfo {title} {{Corner Transfer Matrix Renormalization Group Method}},\ }\href {https://doi.org/10.1143/JPSJ.65.891} {\bibfield  {journal} {\bibinfo  {journal} {Journal of the Physical Society of Japan}\ }\textbf {\bibinfo {volume} {65}},\ \bibinfo {pages} {891} (\bibinfo {year} {1996})}\BibitemShut {NoStop}%
\bibitem [{\citenamefont {Corboz}\ \emph {et~al.}(2010)\citenamefont {Corboz}, \citenamefont {Or\'us}, \citenamefont {Bauer},\ and\ \citenamefont {Vidal}}]{fPEPS}%
  \BibitemOpen
  \bibfield  {author} {\bibinfo {author} {\bibfnamefont {P.}~\bibnamefont {Corboz}}, \bibinfo {author} {\bibfnamefont {R.}~\bibnamefont {Or\'us}}, \bibinfo {author} {\bibfnamefont {B.}~\bibnamefont {Bauer}},\ and\ \bibinfo {author} {\bibfnamefont {G.}~\bibnamefont {Vidal}},\ }\bibfield  {title} {\bibinfo {title} {{Simulation of strongly correlated fermions in two spatial dimensions with fermionic projected entangled-pair states}},\ }\href {https://doi.org/10.1103/PhysRevB.81.165104} {\bibfield  {journal} {\bibinfo  {journal} {Phys. Rev. B}\ }\textbf {\bibinfo {volume} {81}},\ \bibinfo {pages} {165104} (\bibinfo {year} {2010})}\BibitemShut {NoStop}%
\bibitem [{\citenamefont {Jordan}\ \emph {et~al.}(2008)\citenamefont {Jordan}, \citenamefont {Or\'us}, \citenamefont {Vidal}, \citenamefont {Verstraete},\ and\ \citenamefont {Cirac}}]{iePEPS}%
  \BibitemOpen
  \bibfield  {author} {\bibinfo {author} {\bibfnamefont {J.}~\bibnamefont {Jordan}}, \bibinfo {author} {\bibfnamefont {R.}~\bibnamefont {Or\'us}}, \bibinfo {author} {\bibfnamefont {G.}~\bibnamefont {Vidal}}, \bibinfo {author} {\bibfnamefont {F.}~\bibnamefont {Verstraete}},\ and\ \bibinfo {author} {\bibfnamefont {J.~I.}\ \bibnamefont {Cirac}},\ }\bibfield  {title} {\bibinfo {title} {{Classical Simulation of Infinite-Size Quantum Lattice Systems in Two Spatial Dimensions}},\ }\href {https://doi.org/10.1103/PhysRevLett.101.250602} {\bibfield  {journal} {\bibinfo  {journal} {Phys. Rev. Lett.}\ }\textbf {\bibinfo {volume} {101}},\ \bibinfo {pages} {250602} (\bibinfo {year} {2008})}\BibitemShut {NoStop}%
\bibitem [{\citenamefont {Jiang}\ \emph {et~al.}(2008)\citenamefont {Jiang}, \citenamefont {Weng},\ and\ \citenamefont {Xiang}}]{simpleuupdate}%
  \BibitemOpen
  \bibfield  {author} {\bibinfo {author} {\bibfnamefont {H.~C.}\ \bibnamefont {Jiang}}, \bibinfo {author} {\bibfnamefont {Z.~Y.}\ \bibnamefont {Weng}},\ and\ \bibinfo {author} {\bibfnamefont {T.}~\bibnamefont {Xiang}},\ }\bibfield  {title} {\bibinfo {title} {{Accurate Determination of Tensor Network State of Quantum Lattice Models in Two Dimensions}},\ }\href {https://doi.org/10.1103/PhysRevLett.101.090603} {\bibfield  {journal} {\bibinfo  {journal} {Phys. Rev. Lett.}\ }\textbf {\bibinfo {volume} {101}},\ \bibinfo {pages} {090603} (\bibinfo {year} {2008})}\BibitemShut {NoStop}%
\bibitem [{\citenamefont {Corboz}(2016)}]{VPEPS}%
  \BibitemOpen
  \bibfield  {author} {\bibinfo {author} {\bibfnamefont {P.}~\bibnamefont {Corboz}},\ }\bibfield  {title} {\bibinfo {title} {{Variational optimization with infinite projected entangled-pair states}},\ }\href {https://doi.org/10.1103/PhysRevB.94.035133} {\bibfield  {journal} {\bibinfo  {journal} {Phys. Rev. B}\ }\textbf {\bibinfo {volume} {94}},\ \bibinfo {pages} {035133} (\bibinfo {year} {2016})}\BibitemShut {NoStop}%
\bibitem [{\citenamefont {Vanderstraeten}\ \emph {et~al.}(2016)\citenamefont {Vanderstraeten}, \citenamefont {Haegeman}, \citenamefont {Corboz},\ and\ \citenamefont {Verstraete}}]{PhysRevB.94.155123}%
  \BibitemOpen
  \bibfield  {author} {\bibinfo {author} {\bibfnamefont {L.}~\bibnamefont {Vanderstraeten}}, \bibinfo {author} {\bibfnamefont {J.}~\bibnamefont {Haegeman}}, \bibinfo {author} {\bibfnamefont {P.}~\bibnamefont {Corboz}},\ and\ \bibinfo {author} {\bibfnamefont {F.}~\bibnamefont {Verstraete}},\ }\bibfield  {title} {\bibinfo {title} {{Gradient methods for variational optimization of projected entangled-pair states}},\ }\href {https://doi.org/10.1103/PhysRevB.94.155123} {\bibfield  {journal} {\bibinfo  {journal} {Phys. Rev. B}\ }\textbf {\bibinfo {volume} {94}},\ \bibinfo {pages} {155123} (\bibinfo {year} {2016})}\BibitemShut {NoStop}%
\bibitem [{\citenamefont {Liao}\ \emph {et~al.}(2019)\citenamefont {Liao}, \citenamefont {Liu}, \citenamefont {Wang},\ and\ \citenamefont {Xiang}}]{ADPEPS}%
  \BibitemOpen
  \bibfield  {author} {\bibinfo {author} {\bibfnamefont {H.-J.}\ \bibnamefont {Liao}}, \bibinfo {author} {\bibfnamefont {J.-G.}\ \bibnamefont {Liu}}, \bibinfo {author} {\bibfnamefont {L.}~\bibnamefont {Wang}},\ and\ \bibinfo {author} {\bibfnamefont {T.}~\bibnamefont {Xiang}},\ }\bibfield  {title} {\bibinfo {title} {{Differentiable Programming Tensor Networks}},\ }\href {https://doi.org/10.1103/PhysRevX.9.031041} {\bibfield  {journal} {\bibinfo  {journal} {Phys. Rev. X}\ }\textbf {\bibinfo {volume} {9}},\ \bibinfo {pages} {031041} (\bibinfo {year} {2019})}\BibitemShut {NoStop}%
\bibitem [{\citenamefont {Ponsioen}\ \emph {et~al.}(2022)\citenamefont {Ponsioen}, \citenamefont {Assaad},\ and\ \citenamefont {Corboz}}]{AD2}%
  \BibitemOpen
  \bibfield  {author} {\bibinfo {author} {\bibfnamefont {B.}~\bibnamefont {Ponsioen}}, \bibinfo {author} {\bibfnamefont {F.~F.}\ \bibnamefont {Assaad}},\ and\ \bibinfo {author} {\bibfnamefont {P.}~\bibnamefont {Corboz}},\ }\bibfield  {title} {\bibinfo {title} {{Automatic differentiation applied to excitations with projected entangled pair states}},\ }\href {https://doi.org/10.21468/SciPostPhys.12.1.006} {\bibfield  {journal} {\bibinfo  {journal} {SciPost Phys.}\ }\textbf {\bibinfo {volume} {12}},\ \bibinfo {pages} {006} (\bibinfo {year} {2022})}\BibitemShut {NoStop}%
\bibitem [{\citenamefont {Kwan}\ \emph {et~al.}(2021)\citenamefont {Kwan}, \citenamefont {Wagner}, \citenamefont {Soejima}, \citenamefont {Zaletel}, \citenamefont {Simon}, \citenamefont {Parameswaran},\ and\ \citenamefont {Bultinck}}]{IKS}%
  \BibitemOpen
  \bibfield  {author} {\bibinfo {author} {\bibfnamefont {Y.~H.}\ \bibnamefont {Kwan}}, \bibinfo {author} {\bibfnamefont {G.}~\bibnamefont {Wagner}}, \bibinfo {author} {\bibfnamefont {T.}~\bibnamefont {Soejima}}, \bibinfo {author} {\bibfnamefont {M.~P.}\ \bibnamefont {Zaletel}}, \bibinfo {author} {\bibfnamefont {S.~H.}\ \bibnamefont {Simon}}, \bibinfo {author} {\bibfnamefont {S.~A.}\ \bibnamefont {Parameswaran}},\ and\ \bibinfo {author} {\bibfnamefont {N.}~\bibnamefont {Bultinck}},\ }\bibfield  {title} {\bibinfo {title} {{Kekul\'e Spiral Order at All Nonzero Integer Fillings in Twisted Bilayer Graphene}},\ }\href {https://doi.org/10.1103/PhysRevX.11.041063} {\bibfield  {journal} {\bibinfo  {journal} {Phys. Rev. X}\ }\textbf {\bibinfo {volume} {11}},\ \bibinfo {pages} {041063} (\bibinfo {year} {2021})}\BibitemShut {NoStop}%
\bibitem [{sup()}]{suppm}%
  \BibitemOpen
  \href@noop {} {}\bibinfo {note} {See Supplemental Material at XXX for details of numerical methods}\BibitemShut {NoStop}%
\bibitem [{\citenamefont {Lieb}\ \emph {et~al.}(1961)\citenamefont {Lieb}, \citenamefont {Schultz},\ and\ \citenamefont {Mattis}}]{LIEB1961407}%
  \BibitemOpen
  \bibfield  {author} {\bibinfo {author} {\bibfnamefont {E.}~\bibnamefont {Lieb}}, \bibinfo {author} {\bibfnamefont {T.}~\bibnamefont {Schultz}},\ and\ \bibinfo {author} {\bibfnamefont {D.}~\bibnamefont {Mattis}},\ }\bibfield  {title} {\bibinfo {title} {{Two soluble models of an antiferromagnetic chain}},\ }\href {https://doi.org/https://doi.org/10.1016/0003-4916(61)90115-4} {\bibfield  {journal} {\bibinfo  {journal} {Annals of Physics}\ }\textbf {\bibinfo {volume} {16}},\ \bibinfo {pages} {407} (\bibinfo {year} {1961})}\BibitemShut {NoStop}%
\bibitem [{\citenamefont {Oshikawa}(2000)}]{Oshikawa00}%
  \BibitemOpen
  \bibfield  {author} {\bibinfo {author} {\bibfnamefont {M.}~\bibnamefont {Oshikawa}},\ }\bibfield  {title} {\bibinfo {title} {{Commensurability, Excitation Gap, and Topology in Quantum Many-Particle Systems on a Periodic Lattice}},\ }\href {https://doi.org/10.1103/PhysRevLett.84.1535} {\bibfield  {journal} {\bibinfo  {journal} {Phys. Rev. Lett.}\ }\textbf {\bibinfo {volume} {84}},\ \bibinfo {pages} {1535} (\bibinfo {year} {2000})}\BibitemShut {NoStop}%
\bibitem [{\citenamefont {Hastings}(2004)}]{Hastings04}%
  \BibitemOpen
  \bibfield  {author} {\bibinfo {author} {\bibfnamefont {M.~B.}\ \bibnamefont {Hastings}},\ }\bibfield  {title} {\bibinfo {title} {{Lieb-Schultz-Mattis in higher dimensions}},\ }\href {https://doi.org/10.1103/PhysRevB.69.104431} {\bibfield  {journal} {\bibinfo  {journal} {Phys. Rev. B}\ }\textbf {\bibinfo {volume} {69}},\ \bibinfo {pages} {104431} (\bibinfo {year} {2004})}\BibitemShut {NoStop}%
\bibitem [{\citenamefont {Assaad}\ and\ \citenamefont {Herbut}(2013)}]{PhysRevX.3.031010}%
  \BibitemOpen
  \bibfield  {author} {\bibinfo {author} {\bibfnamefont {F.~F.}\ \bibnamefont {Assaad}}\ and\ \bibinfo {author} {\bibfnamefont {I.~F.}\ \bibnamefont {Herbut}},\ }\bibfield  {title} {\bibinfo {title} {{Pinning the Order: The Nature of Quantum Criticality in the Hubbard Model on Honeycomb Lattice}},\ }\href {https://doi.org/10.1103/PhysRevX.3.031010} {\bibfield  {journal} {\bibinfo  {journal} {Phys. Rev. X}\ }\textbf {\bibinfo {volume} {3}},\ \bibinfo {pages} {031010} (\bibinfo {year} {2013})}\BibitemShut {NoStop}%
\bibitem [{\citenamefont {Yang}\ \emph {et~al.}(2021)\citenamefont {Yang}, \citenamefont {Zheng},\ and\ \citenamefont {Qin}}]{PhysRevB.103.155110}%
  \BibitemOpen
  \bibfield  {author} {\bibinfo {author} {\bibfnamefont {X.}~\bibnamefont {Yang}}, \bibinfo {author} {\bibfnamefont {H.}~\bibnamefont {Zheng}},\ and\ \bibinfo {author} {\bibfnamefont {M.}~\bibnamefont {Qin}},\ }\bibfield  {title} {\bibinfo {title} {{Stripe order in the doped Hubbard model on the honeycomb lattice}},\ }\href {https://doi.org/10.1103/PhysRevB.103.155110} {\bibfield  {journal} {\bibinfo  {journal} {Phys. Rev. B}\ }\textbf {\bibinfo {volume} {103}},\ \bibinfo {pages} {155110} (\bibinfo {year} {2021})}\BibitemShut {NoStop}%
\bibitem [{\citenamefont {McCulloch}(2007)}]{McCulloch_2007}%
  \BibitemOpen
  \bibfield  {author} {\bibinfo {author} {\bibfnamefont {I.~P.}\ \bibnamefont {McCulloch}},\ }\bibfield  {title} {\bibinfo {title} {{From density-matrix renormalization group to matrix product states}},\ }\href {https://doi.org/10.1088/1742-5468/2007/10/p10014} {\bibfield  {journal} {\bibinfo  {journal} {Journal of Statistical Mechanics: Theory and Experiment}\ }\textbf {\bibinfo {volume} {2007}},\ \bibinfo {pages} {P10014} (\bibinfo {year} {2007})}\BibitemShut {NoStop}%
\bibitem [{\citenamefont {Rausch}\ \emph {et~al.}(2023)\citenamefont {Rausch}, \citenamefont {Peschke}, \citenamefont {Plorin}, \citenamefont {Schnack},\ and\ \citenamefont {Karrasch}}]{SciPostPhys.14.3.052}%
  \BibitemOpen
  \bibfield  {author} {\bibinfo {author} {\bibfnamefont {R.}~\bibnamefont {Rausch}}, \bibinfo {author} {\bibfnamefont {M.}~\bibnamefont {Peschke}}, \bibinfo {author} {\bibfnamefont {C.}~\bibnamefont {Plorin}}, \bibinfo {author} {\bibfnamefont {J.}~\bibnamefont {Schnack}},\ and\ \bibinfo {author} {\bibfnamefont {C.}~\bibnamefont {Karrasch}},\ }\bibfield  {title} {\bibinfo {title} {{Quantum spin spiral ground state of the ferrimagnetic sawtooth chain}},\ }\href {https://doi.org/10.21468/SciPostPhys.14.3.052} {\bibfield  {journal} {\bibinfo  {journal} {SciPost Phys.}\ }\textbf {\bibinfo {volume} {14}},\ \bibinfo {pages} {052} (\bibinfo {year} {2023})}\BibitemShut {NoStop}%
\bibitem [{\citenamefont {Motruk}\ \emph {et~al.}(2016)\citenamefont {Motruk}, \citenamefont {Zaletel}, \citenamefont {Mong},\ and\ \citenamefont {Pollmann}}]{PhysRevB.93.155139}%
  \BibitemOpen
  \bibfield  {author} {\bibinfo {author} {\bibfnamefont {J.}~\bibnamefont {Motruk}}, \bibinfo {author} {\bibfnamefont {M.~P.}\ \bibnamefont {Zaletel}}, \bibinfo {author} {\bibfnamefont {R.~S.~K.}\ \bibnamefont {Mong}},\ and\ \bibinfo {author} {\bibfnamefont {F.}~\bibnamefont {Pollmann}},\ }\bibfield  {title} {\bibinfo {title} {{Density matrix renormalization group on a cylinder in mixed real and momentum space}},\ }\href {https://doi.org/10.1103/PhysRevB.93.155139} {\bibfield  {journal} {\bibinfo  {journal} {Phys. Rev. B}\ }\textbf {\bibinfo {volume} {93}},\ \bibinfo {pages} {155139} (\bibinfo {year} {2016})}\BibitemShut {NoStop}%
\bibitem [{\citenamefont {Lee}\ \emph {et~al.}(2023)\citenamefont {Lee}, \citenamefont {Sharma}, \citenamefont {Vafek},\ and\ \citenamefont {Changlani}}]{PhysRevB.107.235105}%
  \BibitemOpen
  \bibfield  {author} {\bibinfo {author} {\bibfnamefont {K.}~\bibnamefont {Lee}}, \bibinfo {author} {\bibfnamefont {P.}~\bibnamefont {Sharma}}, \bibinfo {author} {\bibfnamefont {O.}~\bibnamefont {Vafek}},\ and\ \bibinfo {author} {\bibfnamefont {H.~J.}\ \bibnamefont {Changlani}},\ }\bibfield  {title} {\bibinfo {title} {{Triangular lattice Hubbard model physics at intermediate temperatures}},\ }\href {https://doi.org/10.1103/PhysRevB.107.235105} {\bibfield  {journal} {\bibinfo  {journal} {Phys. Rev. B}\ }\textbf {\bibinfo {volume} {107}},\ \bibinfo {pages} {235105} (\bibinfo {year} {2023})}\BibitemShut {NoStop}%
\bibitem [{\citenamefont {Morera}\ \emph {et~al.}(2023)\citenamefont {Morera}, \citenamefont {Kan\'asz-Nagy}, \citenamefont {Smolenski}, \citenamefont {Ciorciaro}, \citenamefont {Imamo\ifmmode~\breve{g}\else \u{g}\fi{}lu},\ and\ \citenamefont {Demler}}]{morera2023high}%
  \BibitemOpen
  \bibfield  {author} {\bibinfo {author} {\bibfnamefont {I.}~\bibnamefont {Morera}}, \bibinfo {author} {\bibfnamefont {M.}~\bibnamefont {Kan\'asz-Nagy}}, \bibinfo {author} {\bibfnamefont {T.}~\bibnamefont {Smolenski}}, \bibinfo {author} {\bibfnamefont {L.}~\bibnamefont {Ciorciaro}}, \bibinfo {author} {\bibfnamefont {A.~m.~c.}\ \bibnamefont {Imamo\ifmmode~\breve{g}\else \u{g}\fi{}lu}},\ and\ \bibinfo {author} {\bibfnamefont {E.}~\bibnamefont {Demler}},\ }\bibfield  {title} {\bibinfo {title} {{High-temperature kinetic magnetism in triangular lattices}},\ }\href {https://doi.org/10.1103/PhysRevResearch.5.L022048} {\bibfield  {journal} {\bibinfo  {journal} {Phys. Rev. Res.}\ }\textbf {\bibinfo {volume} {5}},\ \bibinfo {pages} {L022048} (\bibinfo {year} {2023})}\BibitemShut {NoStop}%
\bibitem [{\citenamefont {Chen}\ \emph {et~al.}(2024)\citenamefont {Chen}, \citenamefont {Jiang}, \citenamefont {Zhang},\ and\ \citenamefont {Hu}}]{Chen2024}%
  \BibitemOpen
  \bibfield  {author} {\bibinfo {author} {\bibfnamefont {Y.}~\bibnamefont {Chen}}, \bibinfo {author} {\bibfnamefont {K.}~\bibnamefont {Jiang}}, \bibinfo {author} {\bibfnamefont {Y.}~\bibnamefont {Zhang}},\ and\ \bibinfo {author} {\bibfnamefont {J.}~\bibnamefont {Hu}},\ }\bibfield  {title} {\bibinfo {title} {{Flux phases in the extended Hubbard model on the triangular lattice}},\ }\href {https://doi.org/10.1007/s11433-024-2413-5} {\bibfield  {journal} {\bibinfo  {journal} {Sci. China Phys., Mech. {\&} Astron.}\ }\textbf {\bibinfo {volume} {67}},\ \bibinfo {pages} {297211} (\bibinfo {year} {2024})}\BibitemShut {NoStop}%
\end{thebibliography}%

\renewcommand{\theequation}{S\arabic{equation}}
\setcounter{equation}{0}
\renewcommand{\thefigure}{S\arabic{figure}}
\setcounter{figure}{0}
\renewcommand{\thetable}{S\arabic{table}}
\setcounter{table}{0}

%%%%%%%%%%%
\subsection{Details of Hartree-Fock calculation}
We first determine quantum phases within the Hartree-Fock approximation. Compared to previous studies, we adopt a less constrained approach: we conserve the total particle number and scan the ordering momenta $\mathbf{Q}$ over grids up to size $48\times48$ with no further restrictions. The ``unrestricted" search is performed with random initial states followed by ODA self-consistent optimization~\cite{ODA}. The dimensions of the ordering momentum grids are factors of the total momentum grid size. For example, $144 \times 144$ for $48\times48$ $\mathbf{Q}$ mesh,  $125 \times 125$ for $5 \times 5$ $\mathbf{Q}$ mesh. This commensuration constraint can be avoided for the single-$\mathbf{Q}$ (canted) spin spiral states $\langle S^{+}(i) \rangle =m^{(+)} e^{i \mathbf{Q}\cdot\mathbf{r}_i},$ $\langle S^{z}(i) \rangle =  0 \ (m_z) $ through a ``co-rotating-frame"  transformation, converting them to uniform states. We thus additionally consider incommensurate $\mathbf{Q}$ for these spiral states. For all calculations, we enlarge the momentum grid until finite-size energy errors are below $10^{-5}$ per site. For our unrestricted calculations, we have compared converged results obtained from several different random initial states, with optimized energies differing by much less than $10^{-5}$. 
This precision is required because competing HF states can differ in energy by only $10^{-4}$ per site. (We target the 3QII state using random collinear initial states and find the difference $\sim 10^{-4}$ with 3QI states.)  We note that while we pursue $10^{-5}$ energy accuracy within HF, the HF approximation itself usually induces much larger deviations from the exact results.  
For the Hubbard model, HF can give a rigorous bound for the existence of a fully-polarized (FP) ferromagnetic ground state, based on the fact that HF energy is exact for this state. Thus, our HF results give a lower bound for $U$ below which there is no FP. 

\subsection{Details and additional data of density matrix renormalization group  study}

\begin{figure}
\begin{center}
\includegraphics[width=1\columnwidth]{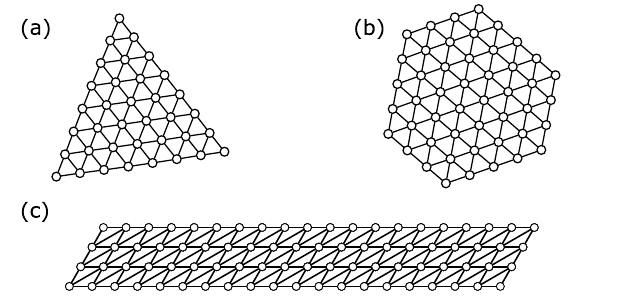}
\end{center}
\caption{
Finite triangular lattices studied by DMRG. (a) 36-site triangular flake. (b) 48-site hexagonal flake. (c) Four-leg zigzag cylinder with length 20 (periodic boundary condition imposed for the tangential direction).
} 
\label{fig:dmrglattices}
\end{figure}

\begin{figure}
\begin{center}
\includegraphics[width=0.7\columnwidth]{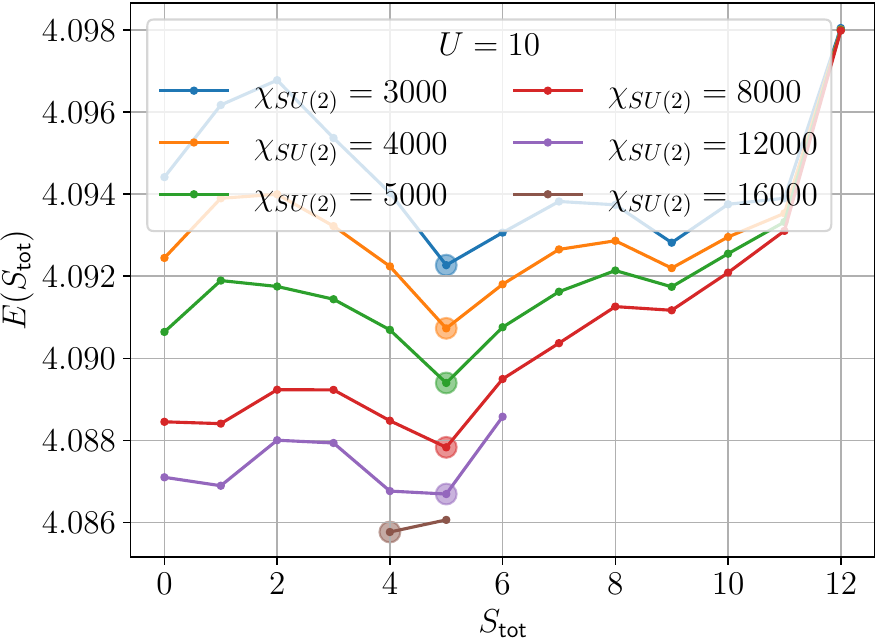}
\end{center}
\caption{
DMRG data example for ferromagnetization for the 48-site hexagonal flake. The plot shows the lowest energy in a given sector of the total spin $S_{\mathrm{tot}}$ at $U=10$ for various SU(2)-invariant bond dimensions. The largest energy errors are estimated to be in the order $10^{-3}$. The triangular and cylindrical geometry has one and two orders of smaller errors. 
} 
\label{fig:Stotdata}
\end{figure}

Our density matrix renormalization group (DMRG) calculations are performed for flakes and cylinder geometries illustrated in Fig~\ref{fig:dmrglattices}. We implement the charge U$(1)$ symmetry to fix the density and the spin SU$(2)$ to control  $S_{\mathrm{tot}}$ quantum numbers. For cylinder geometries, we additionally impose transverse momentum quantum numbers. We compare the energies of the lowest-energy states of different quantum number sectors to determine the $S_{\mathrm{tot}}$ quantum number of the ground state.  This provides not only greater numerical efficiency but also a stronger validation of convergence, since all $S_{\mathrm{tot}}$ are checked. As an example, in Fig.~\ref{fig:Stotdata}, we plot a set of data for convergence validation. In the main text, we plot the polarization per site of the highest-weight state, excluding the edge sites. This is because we find that there is a range of $U$ where the ground state is in the sector of 1/2 less than the maximal $S_{\mathrm{tot}}$ and the deviation of $S_z$ from the maximum (0.25) is distributed mainly along the edge. So we consider that it is better to exclude the edge sites to determine the full polarization of the bulk. 

The spontaneous SDW orders cannot be detected directly in finite systems. We infer it from the static structural factor calculated in the cylindrical geometry [Fig.1(b)]:
%[Fig.~\ref{fig:TN32}b]: 
$S(\bm{k})=\sum_{m}e^{i \bm{k}\cdot (\bm{r}_m-\bm{r}_n)}[\langle \bm{S}(m) \cdot \bm{S}(n)\rangle-\langle \bm{S}(m)\rangle \cdot \langle \bm{S}(n)\rangle]$ averaged for the central sites $m$ of the cylinder geometry. Strictly speaking, the static structure factor is defined for periodic boundary conditions as $S(\bm{k})=\sum_{m}e^{i \bm{k}\cdot (\bm{r}_m-\bm{r}_n)}[\langle \bm{S}(m) \cdot \bm{S}(n)\rangle-\langle \bm{S}(m)\rangle \cdot \langle \bm{S}(n)\rangle]$.   We use an open boundary condition for the axial direction in DMRG, because otherwise DMRG is less efficient. As $m$ is chosen to be far from the boundaries, the result should be a good proxy for the results of the period boundary condition. The ordering momentum $\bm{Q}$ of the SDWs can be read as peaks of $S(\bm{k})$. In 
%Fig.~\ref{fig:TN32}(b), 
Fig.1(b), 
the results of $U=8$ in $4 \times 20 $ there is only one pair of obvious peaks, located at one $M-\Gamma$ line at some apparently incommensurate value. The unidirectional correlation could be an indication of a spiral or collinear stripe phase.  Since the correlation function is SU$(2)$ symmetric, it is not obvious how to distinguish coplanar and collinear phases. We also notice from 
%[Fig.~\ref{fig:TN32}(a)] 
[Fig.1(a)]
that the state has a non-zero net polarization. Thus, a canted spiral phase is the most likely candidate. A canted spiral has not been found as the ground state within the HF approximation, but it is closely competing with the optimal HF ground state. As mentioned in the main text, we do not find evidence for other SDWs upon tuning $U$. For example, for 3Q orders, the static structure factor should peak at $(0, \pi)$, $(\pi, 0)$, and $(\pi, \pi)$, but these peaks are missing in our data of $4 \times 20 $ systems. We have argued that this is a finite-size effect as these orders are also absent in similar-sized HF calculations. Here, we provide an explanation of the finite-size effects of the cylinder geometry by analyzing the Fermi surface. The infinitesimal interaction instability that drives density-wave formation requires Fermi surface nesting with the 3Q nesting wave vector.
It is straightforward to verify that the mini bands of the four-leg zigzag cylinder have quasi-1D nesting momentum other than 3Q.

The polarization curve of $4 \times 20 $ is different from those of two flakes for intermediate interaction: the polarization becomes non-zero at a smaller $U$. This might be an artifact of the geometry. The four-leg zigzag cylinder has a band structure with one of the four 1D bands (labeled by different $k_y$ momenta, with the $y$-direction being the compact direction) being completely flat. Although a single flat 1D band can exist for larger zigzag cylinders, the number of dispersive bands increases. We therefore speculate that the 1D flat band is responsible for the large polarization at intermediate $U$, which might not persist to the two-dimensional limit.

\subsection{Details of variational infinite projected entangled pair states  study}

\begin{figure}
\begin{center}
\includegraphics[width=0.5\columnwidth]{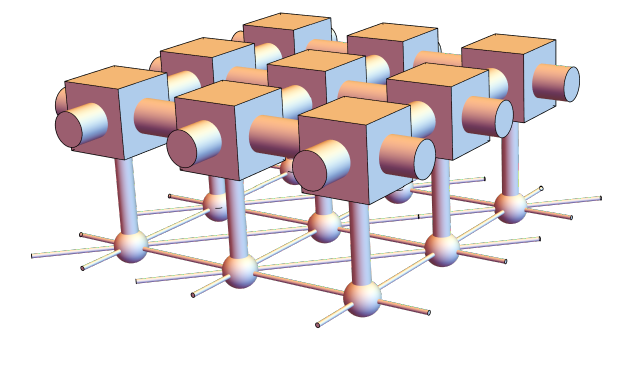}
\end{center}
\caption{Implemented PEPS tensor network. The cubes in the top layer represent tensors. 
Four bonds of each tensor connect to the neighboring tensors; each bond corresponds to an index of the tensor with dimension D. The remaining fifth bond of a tensor carries the local Hilbert space of a site, represented by a sphere. The bonds between the spheres represent couplings of the Hamiltonian on the triangular lattice.  The network is infinite, and a part of it is shown here.
} 
\label{fig:PEPSnet}
\end{figure}

We adopt square-lattice-like projected entangled pair states (PEPS), i.e., for the tensor at each site, there is one leg for the local Hilbert space (dimension 4) and four legs  (dimension $D$) connecting four out of six physical nearest neighbors (Fig.~\ref{fig:PEPSnet}). The physical neighbor structure of the Hamiltonian is restored by coupling not only four nearest neighbors on the square-lattice-like network but also 
two out of four next-nearest neighbors (diagonals).
This technical choice simplifies the implementation, since it enables one to reuse the toolbox developed for the square lattice, though it comes with the drawback that the lattice symmetries are not reproduced exactly at low D, but they are typically restored at sufficiently large D.

We implement  charge U(1) conservation to fix the charge density and to accelerate calculations. This puts a restriction on the choice of the unit cell: The number of particles in each unit cell must be an integer. For example, the choice $3\times 3$ for $3/2$ is not possible because the number of particles per unit cell, 27/2, is not an integer. Therefore, the unit cell has to be enlarged to at least $3\times 6$. We have implemented PEPS states with both real and complex numbers. Using real tensors restricts the magnetic order to the xz plane. However, even with complex tensors we do not obtain clear evidence of non-coplanar states or finite chirality.

The calculation of physical quantities, e.g., energy densities and magnetic moments, and also the gradient for optimizations, corresponds to contracting the infinite PEPS. 
We adopt the corner transfer matrix (CTM) method~\cite{CTM} for the approximate contraction, where the contraction error is controlled by an additional bond dimension $\chi$. In our calculations, we use $3D^2\leq \chi \leq 9D^2$ for which the contraction errors of the measurements are much smaller than the symbol sizes. For optimization, we typically use $\chi=4D^2$. 

To optimize the PEPS, we implement a gradient descent method (Broyden-Fletcher-Goldfarb-Shanno algorithm) using automatic differentiation. This method has been demonstrated to be the state-of-the-art in previous studies of spin models~\cite{VPEPS,ADPEPS}, compared to traditional methods based on imaginary time evolution with simple and full update optimization.  We 
find that in our case of a fermionic system with frustration, the simple update method may be too inaccurate to target the ground states, while gradient descent provides the highest accuracy.

Here, we comment on the finite $D$ effects of PEPS,  which helps to clarify the bias in the results for inferring quantum phases. The bias arises because finite-$D$ energy errors are different for different phases. The correct quantum phases may be missed due to large finite-$D$ errors. In our cases, all competing physical states have magnetic orders and are thus gapless in the spin sector. The 3QI state is a Chern insulator, while other phases are metallic. Both cases have larger finite-D errors than trivial insulators. A priori, it is not clear which of these phases have larger finite D errors.
Nevertheless, $D=1$ PEPS can not represent a tetrahedral order for $\overline{\langle n_i \rangle}=\frac{3}{2}$ but it can represent the collinear order we found for all $D \leq 8$. Thus, it is reasonable to conjecture that with relatively small $D$, the variational energy errors for tetrahedral states are large. Even our collinear states ($D=8$) appear to have variational energy errors greater than 0.005, which is much larger than the competing energy scale suggested by HF ($\sim 10^{-4}$). Thus, we conclude that our PEPS calculation cannot rule out the existence of tetrahedral ground states.

Finite-$D$ PEPS can also distort the order parameters. When targeting a particular gapless ground state, the best finite-$D$ PEPS is typically a gapped state with the correlation length increasing with increasing $D$. The symmetries that enforce the gaplessness due to an LSM theorem thus cannot be exactly realized by finite-$D$ PEPS. As mentioned in the main text, the spiral-like state we obtained for $U=9$ has an additional amplitude modulation of the magnetic moment. Such an amplitude modulation could be a finite-$D$ effect that results in an artificial gap opening, as this modulation breaks the generalized translation symmetry, which (together with charge conservation) enforces the gaplessness due to an LSM theorem. A similar situation may occur for the 3Q-II state, which may also be distorted in finite-$D$ PEPS. In our case, we choose a doubled PEPS unit cell $3 \times 6$), but it is unclear if there is a tendency to restore the $3 \times 3$ unit cell when increasing $D$.

\begin{figure}
\begin{center}
\includegraphics[width=0.8\columnwidth]{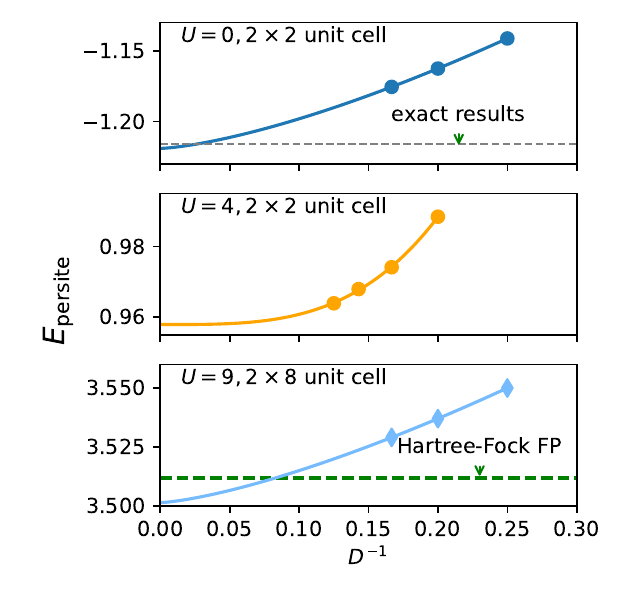}
\end{center}
\caption{Finite-D extrapolation for energies. The extrapolation ansatz is $E(D)=E_0+ c/D^\alpha$. The Hartree-Fock results are referenced as dashed lines where for $U=0$ it is exact. For $U=4$, iPEPS energy is well below  the HF energy [main text Fig.2(c)] which is not shown here.}
\label{fig:PEPSenergyextrapolation}
\end{figure}

To have a better estimate of energy competitions among quantum phases, we perform finite $D$ extrapolation.  To test the method, we also consider the free fermion case ($U=0$). Examples of extrapolation are shown in Fig.~\ref{fig:PEPSenergyextrapolation}. For $U=0$, we compare the iPEPS results with the numerically exact results obtained from band calculations. The errors of the extrapolated results are reduced by more than $90\%$ compared to the  $D=6$ results.   We expect that the $U=0$ calculations are the most challenging because they target spinful Fermi liquids with high entanglement. We use $\chi=7D^2$ in optimization, which makes it expensive to scale up $D$.  For finite $U$, we can estimate the finite energy errors $D$ by comparing with the extrapolated results, which are significantly lower than that of $U=0$.  For $U=4$, the iPEPS energy is significantly lower than Hartree-Fock. For $U=9$, Hartree-Fock energy competes. We interpret the reason as the competing energy scale between spiral and fully polarized (FP) is small, for the latter Hartree-Fock is exact. The extrapolation helps infer that spiral states are favored.

\end{document}